\RequirePackage{rotating}

\documentclass[12pt,a4paper,fleqn]{article}
\usepackage[utf8]{inputenc}
\usepackage{authblk}
\usepackage{geometry}
\usepackage{xcolor}
\usepackage[round]{natbib}
\usepackage{graphicx}
\usepackage{pdflscape}  
\usepackage{appendix}
\usepackage{threeparttable}
\usepackage{booktabs}
\usepackage{subcaption}
\usepackage{longtable}
\usepackage{threeparttable}

\usepackage{amssymb}
\usepackage{amsbsy}
\usepackage{bm}
\usepackage{amsmath}
\usepackage{amsthm}
\usepackage{graphicx}
\usepackage{mathtools}
\usepackage{setspace}
\usepackage[justification=centering]{caption}

\usepackage{color, colortbl}
\definecolor{ashgrey}{rgb}{0.7, 0.75, 0.71}
\definecolor{columbiablue}{rgb}{0.61, 0.87, 1.0}
\definecolor{coral}{rgb}{1.0, 0.5, 0.31}

\definecolor{colBVAR}{HTML}{bababa}
\definecolor{colBART}{HTML}{d7191c}
\definecolor{colmixBART}{HTML}{fdae61}
\definecolor{colerrorBART}{HTML}{abd9e9}
\definecolor{colfullBART}{HTML}{2c7bb6}

\definecolor{colcons}{HTML}{e31a1c}
\definecolor{colSV}{HTML}{a6cee3}
\definecolor{colhBART}{HTML}{1f78b4}

\usepackage{enumitem}
\newlist{steps}{enumerate}{1}
\setlist[steps,1]{label = Step \arabic*:}

\usepackage{adjustbox}
\usepackage{dcolumn}
\newcolumntype{d}[1]{D..{#1}} 

\usepackage{xcolor,colortbl}

\definecolor{Gray}{gray}{0.85}
\definecolor{LightCyan}{rgb}{0.88,1,1}

\newcolumntype{a}{>{\columncolor{Gray}}c}
\newcolumntype{b}{>{\columncolor{white}}c}
\usepackage{array}
\newcolumntype{H}{>{\setbox0=\hbox\bgroup}c<{\egroup}@{}}
\newcolumntype{Z}{>{\setbox0=\hbox\bgroup}c<{\egroup}@{\hspace*{-\tabcolsep}}}

\usepackage{caption}
\usepackage{subcaption}
\definecolor{nblue}{HTML}{000660}
\usepackage[colorlinks=true,urlcolor=nblue,linkcolor=nblue,citecolor=nblue]{hyperref}
\newcommand*{\myeqref}[2][Eq.~]{%
  \hyperref[{#2}]{#1(\ref*{#2})}%
}
\def\equationautorefname#1#2\null{%
  Eq.#1(#2\null)%
}

\begin{document}
\title{\textbf{Fast and Order-invariant Inference in Bayesian VARs with Non-Parametric Shocks}\thanks{
 Huber gratefully acknowledges financial support from the Austrian Science Fund (FWF, grant no. ZK 35). }}

\author[a,c]{Florian \textsc{Huber}}
\author[b]{Gary \textsc{Koop}}
\affil[a]{\textit{University of Salzburg}}
\affil[b]{\textit{University of Strathclyde}}
\affil[c]{\textit{International Institute for Applied Systems Analysis (IIASA)}}
\date{\today}

\maketitle\thispagestyle{empty}\normalsize\vspace*{-2em}\small\linespread{1.5}
\begin{center}
\begin{minipage}{0.8\textwidth}

\begin{center}
    \small \textbf{Abstract}
\end{center}
\noindent  The shocks which hit macroeconomic models such as Vector Autoregressions (VARs) have the potential to be non-Gaussian, exhibiting asymmetries and fat tails. This consideration motivates the VAR developed in this paper which uses a Dirichlet process mixture (DPM) to model the shocks. However, we do not follow the obvious strategy of simply modeling the VAR errors with a DPM since this would lead to computationally infeasible Bayesian inference in larger VARs and potentially a sensitivity to the way the variables are ordered in the VAR. Instead we develop a particular additive error structure inspired by Bayesian nonparametric treatments of random effects in panel data models. We show that this leads to a model which allows for computationally fast and order-invariant inference in large VARs with nonparametric shocks. Our empirical results with nonparametric VARs of various dimensions shows that nonparametric treatment of the VAR errors is particularly useful in periods such as the financial crisis and the pandemic.
\vspace{-0.1cm}

\end{minipage}
\end{center}

\normalsize\newpage
\section{Introduction}
Bayesian Vector Autoregressions (VARs) are now routinely used with large numbers of dependent variables. Use of non-conjugate priors or non-Gaussian error distributions typically requires the use of Markov Chain Monte Carlo (MCMC) methods, which leads to a large computational burden. This means full system estimation of the reduced form VAR is difficult or infeasible with large VARs. This has led many researchers to avoid full system estimation and instead work with a structural VAR with a diagonal error covariance matrix. The structural VAR allows for estimating one equation at a time which greatly reduces the computational burden making Bayesian estimation of large VARs practical. However, standard specifications for the structural VAR  which allow for equation-by-equation estimation (e.g. \cite{CCM2019}) suffer from order dependence (i.e. posterior and predictive densities depend on the manner in which the variables are ordered in the VAR). The importance of order dependence, and in particular its impact on predictive variances in larger VARs, is discussed in papers such as \cite{ARRS21} and  \cite{chan2021large}. There have been some order invariant approaches proposed which do allow for equation-by-equation estimation, including \cite{chan2021large} and \cite{wu_eigen} but these assume Gaussian errors and the former relies on the presence of stochastic volatility to identify the model. However, the presence of large VAR shocks which imply sudden shifts in variances and/or asymmetries in predictive densities means more flexibility is required. These considerations motivate the present paper where we develop a VAR with a nonparametric non-Gaussian error distribution for the shocks. The MCMC algorithm we derive for this VAR is computationally efficient and order invariant. 

There is a growing VAR literature which wishes to develop flexible models for the VAR errors. However, our model differs from this literature by its use of non-parametric methods and its focus on computational efficiency and order invariance. For instance, several papers, including \cite{CHIU2017} and \cite{KARLSSON2023104580} work with VARs with parametric, non-Gaussian error distributions (e.g. Student-t distributions), but do not allow for equation-by-equation estimation and, thus, are computationally slower than our approach. A VAR with a non-parametric distribution for the shocks is \cite{RePEc:boe:boeewp:0957} which does allow for equation-by-equation estimation. But the flexible error distribution in this case is used to identify structural shocks and does not address the ordering issue. And, in addition, the MCMC algorithm becomes more involved due to the fact that the contemporaneous relations need to be sampled using Metropolis-Hastings (MH) updates.

To develop our model, we borrow ideas from the literature on semi- and non-parametric estimation of random effects in panel data models, see \cite{FSTO} or \cite{Dunson}. The key insight is that we can exploit the random effect representation of the covariance matrix of the system to enable equation-by-equation estimation, see \cite{FoxDunson}. We decompose the shock vector of the VAR into two components. The first component is a vector of random effects that feature an unknown multivariate distribution which exhibits correlation between the errors in different equations. The second component is a vector of Gaussian random shocks which are uncorrelated across equations. Conditional on the random effects, the model becomes a system of uncorrelated regression models. But after integrating out the random effects, the resulting shock distribution features cross-sectional dependence. The key implication is that fast estimation is possible since the VAR coefficients can be drawn one equation at a time conditional on the random effects. By assuming a Dirichlet Process Mixture (DPM) for the vector of random effects we achieve great flexibility since the joint distribution of the shocks can be skewed, feature heavy tails, be heteroskedastic or be multi-modal.  In the multiple equation VAR context, this flexibility is potentially of great benefit since it allows for the errors in the different equations to have different properties. 

In an exercise using artificial data, we show how our non-parametric VAR can automatically uncover a variety of departures from Gaussianity. 
In an empirical exercise involving a large data set of US macroeconomic variables, we demonstrate the advantages of our model both for forecasting and for structural economic analysis.

The remainder of the paper is structured as follows. The next section shows how a random effects representation of the VAR can be used to facilitate equation-by-equation estimation and how we treat non-parametric shocks in the VAR. Section \ref{sec: bayes} discusses the prior setup, sketches the MCMC algorithm and discusses some computational details. Sections \ref{sec: simulation} and \ref{sec: forecasting} apply the model to synthetic and real data, respectively. The final section gives a summary and concludes the paper.

\section{VARs with non-parametric shocks}\label{sec: econometrics}
\subsection{A linear VAR with an additive Gaussian error structure}\label{sec: linear}
Before introducing our non-parametric specification for the shocks to the VAR, it is instructive to begin with a parametric version of our model. This involves an $M$-dimensional vector of dependent variables, $\{\bm y_t\}_{t=1}^T$, which evolves as:
\begin{equation}
    \bm y_t = \bm A \bm X_t + \bm \epsilon_t +  \bm v_t, \quad 
    \bm \epsilon_t \sim \mathcal{N}(\bm 0_M, \bm  \Sigma) \quad  \bm v_t \sim \mathcal{N}(\bm 0_M, \bm  \Omega_t)
    \label{VAR}
\end{equation}
where $\bm X_t = (\bm y'_{t-1}, \dots, \bm y'_{t-p})'$ is a $K(=Mp)$ matrix of lagged endogenous variables 
and $\bm A$ denotes an $M \times K$ matrix of VAR coefficients. 
This specification differs from a conventional VAR in that it has two $M$-dimensional errors, $\bm \epsilon_t$ and $\bm v_t$, which are assumed to be independent over time and of one another at all leads and lags. The only restriction on $\bm \Sigma$ is that it is positive definite whereas $\bm \Omega_t$ is restricted to be a diagonal matrix with individual error variances $\omega_{1t}, \dots, \omega_{Mt}$.  We assume that the logarithms of $\omega_{jt}$ evolve according to AR(1) processes leading to a standard SV specification.
The covariance matrix of the VAR errors, $\bm \varepsilon_t = \bm \epsilon_t + \bm v_t$ is $\bm \Xi_t = \bm \Sigma + \bm \Omega_t$. Notice that the SV assumption on the idiosyncratic shocks implies that the main diagonal elements of $\bm \Xi_t$ are given by $\sigma_{ii}^2 + \omega_{it}$, where $\sigma_{ii}^2$ is the $(i, i)^{th}$ element of $\bm \Sigma$.\footnote{We note that, without further restrictions, this specification is not identified. But this lack of identification poses no problem for Bayesian estimation and prediction with this model provided a proper prior is used. Furthermore, if the researcher wishes to impose identification, this can easily be done in standard cases. For instance, beginning with  \cite{COGLEY2005262} many popular Bayesian VARs have assumed the $\omega_{it}$ to follow stochastic volatility (SV) processes. Fixing the SV initial conditions to be zero suffices to identify the model.} 

The reasons why we adopt this additive error structure relate to efficient computation and order invariance. To explain these points, we first summarise some key issues in the Bayesian VAR literature which are 
particularly acute with larger VARs. The traditional reduced form VAR is given by (\ref{VAR}) but with $\bm v_t$ set to zero.\footnote{Typically, the assumption of homoskedasticity is relaxed, but allowing for this does not affect the arguments in this sub-section.} Direct Bayesian estimation of this reduced form VAR is computationally challenging when $M$ is large due to the need to carry out matrix manipulations involving the very high dimensional posterior covariance matrix of $\bm A$. Accordingly, it is common to carry out Bayesian estimation in a structural form involving the use of the Cholesky decomposition of the reduced form error covariance matrix. That is, decomposing $\bm \Sigma = \bm B \bm D \bm B^{'} $ where $\bm B$ is lower triangular with ones on the diagonal and $\bm D$ is diagonal, the structural VAR is obtained by multiplying both sides of the VAR by $\bm B^{-1}$. The structural VAR has a diagonal error covariance matrix which means estimation can be carried out one equation at a time. This leads to huge computational improvements. For instance, in the specification used in \cite{CCM2019} the MCMC algorithm based on the reduced form VAR requires $ O(M^6)$ elementary operations to take one draw of the VAR coefficients but only $O(M^4)$ with the structural VAR. Thus, there are enormous computational benefits from working with VARs specified in such a way as to allow equation-by-equation estimation. 

However, Bayesian results using the structural VAR based on the Cholesky decomposition are order dependent (i.e. posterior and predictive results depend on the way the variables are ordered in the VAR). This contrasts with the reduced form VAR for which standard implementations (e.g. use of an inverse Wishart prior for $\bm \Sigma$) are order invariant. The empirical importance of ordering issues has been investigated in papers such as \cite{ARRS21} and  \cite{chan2021large} and found to be substantive, particularly in the case of large VARs and particularly for predictive variances and higher order predictive moments. Thus, most Bayesian VAR papers either work in reduced form, and face computational challenges unless the VAR dimension is very low, or work in structural form and produce empirical results which depend on the way the variables are ordered in the VAR.

If we now return to our VAR with additive errors in (\ref{VAR}), it is straightforward to show that it suffers neither of these drawbacks. Computationally efficient MCMC algorithms can be developed which exploit the fact that (conditional on $\bm \epsilon_t$) the equations are independent of one another. But marginally (i.e. after integrating out $\bm \epsilon_t$), the shocks to $\bm y_t$ are cross-sectionally correlated. In relation to order-invariance, it is worth emphasizing that the ordering issue does not relate to the likelihood (i.e. the structural and reduced form VARs lead to the same likelihood function) but rather relates to the prior that is placed on the error variances and covariances. Indeed \cite{CCM2019} refer to it as the "prior ordering issue". In our additive error setup, order invariance can be achieved by retaining an inverse Wishart prior for $\bm \Sigma$. Since $\bm \Omega_t$ is diagonal any conventional set of priors will lead to order invariance. For instance, in the homoskedastic version of the model, assuming $\omega_{i}$ for $i=1 \ldots M$ to have inverse-Gamma priors which are independent across $i$ leads to order invariance. In the heteroskedastic case, assuming $\omega_{it}$ to have independent SV processes leads to order invariance. 

In subsequent sections, we will work with a non-parametric version of this model and provide full details of the priors we use and our computationally efficient MCMC algorithm. We stress that the issues discussed in this sub-section also hold with non-parametric VARs. However, this subsection  provides the basic insights into how these computational benefits are achieved and why our prior is order invariant. In addition, it may be found useful by Bayesian VAR researchers who are happy to remain parametric and work with linear VARs with Gaussian shocks. Specifying the VAR as we have done, with additive errors, is an attractive way of achieving fast, order invariant inference even in conventional VARs.

\subsection{Allowing for shocks of unknown form}\label{sec: likelihood}
The model in the preceding subsection had a linear conditional mean and Gaussian error structure. In many contexts, linearity and Gaussianity can be restrictive and this is particularly so in extreme times such as the recent Covid-19 pandemic. In this paper we focus on relaxing the Gaussianity assumption relating to the shocks hitting the model. We will maintain the assumption of a linear conditional mean and assume a standard SV process for $\bm \Omega_t$.\footnote{Adopting non-parametric approaches for either of these can easily be done. For instance, \cite{ huber2020inference}, \cite{huber2020nowcasting} and \cite{chkmp2021tail} model the conditional mean of a VAR non-parametrically using regression trees. The last of these papers also uses regression trees to model the conditional variance. Approaches such as these could be added to the model of the present paper if extra flexibility is desired. However, as will be demonstrated below our model is already very flexible and can model any of the empirical regularities common with macroeconomic data.} In particular our non-parametric VAR is the same as the one specified in the preceding section except for assumptions relating to $\bm \epsilon_t$. The assumption of Gaussianity will be replaced by a Dirichlet process mixture (DPM) of Gaussians. We will show that by this simple extension, we will achieve great gains in empirically-relevant flexibility. But because the error process remains conditionally Gaussian, the benefits discussed in the preceding sub-section (i.e. equation-by-equation estimation and order invariance) will be retained. 

The ideas underlying our treatment of $\bm \epsilon_t$ are inspired by papers such as \cite{FSTO} and \cite{Dunson} which develop parametric and non-parametric Bayesian treatments of random effects in panel data models and, accordingly, we refer to $\bm \epsilon_t$ as  a vector of random effects. 
We model the random effects by introducing a base measure $\mathcal{G}$ and defining a parametric family of component densities $f$ with unknown parameters $\bm \vartheta$:
 \begin{equation*}
        p(\bm \epsilon_t) = \int f(\bm \epsilon_t | \bm \vartheta) ~\mathcal{G}(d \bm \vartheta) = \sum_{j=1}^\infty \eta_j f(\bm \epsilon_t | \bm \vartheta_j),
\end{equation*}
where the weights $\sum_{j=1}^\infty \eta_j$ sum to $1$. We assume that the component densities $f$ are Gaussian with $M \times 1$ mean vector $\bm \mu_j$ and $M \times M$ variance-covariance matrix $\bm \Sigma_j$
which implies 
\begin{equation*}
    p(\bm \epsilon_t) = \sum_{j=1}^\infty \eta_j \mathcal{N}(\bm \mu_j, \bm \Sigma_j).
\end{equation*}
Alternatively, this model can be written in terms of a discrete latent random variable $\delta_t \in \{1,2 \dots \}$ that indicates which mixture component to adopt in time $t$:
    \begin{equation}
        \bm \epsilon_t = \bm \mu_{\delta_t} + \bm Q_{\delta_t} \bm w_t, \quad \bm w_t \sim \mathcal{N}(\bm 0, \bm I_M)
        \label{efactor}
    \end{equation}
with $\text{Prob}(\delta_t = j) = \eta_j$  and $\bm Q_{\delta_t}$ denoting the lower Cholesky factor of $\bm \Sigma_{\delta_t}$ with $\bm \Sigma_{\delta_t} = \Sigma_j$ if $\delta_t = j$.\footnote{Note that this use of the Cholesky decomposition does not undermine the  order invariance of the model since we will use an inverse Wishart prior for $\bm \Sigma_j$ which is order-invariant.} 
Plugging this expression for $\bm \epsilon_t$ into (\ref{VAR}) gives
\begin{equation}
    \label{BNPVAR}
        \bm y_t = \bm \mu_{\delta_t} + \bm A \bm X_t + \underbrace{\bm Q_{\delta_t} \bm w_t + \bm v_t}_{\bm \varepsilon_t}
\end{equation}
 which is VAR with reduced-form shocks $\bm \varepsilon_t \sim \mathcal{N}(\bm 0_M, \bm \Xi_t)$ and time-varying error covariance matrix $\bm \Xi_t = \bm \Sigma_{\delta_t} + \bm \Omega_t$. Another interesting interpretation of $\bm \varepsilon_t$ is that it resembles a factor model with $M$ factors and a particular form of the factor loadings.  

The model given in (\ref{BNPVAR}) has several properties which make it attractive for use in empirical macroeconomics. It retains a conditionally (i.e. conditional on $\delta_t$) Gaussian structure which leads to simplicity of computation and structural economic interpretation. However, it is extremely flexible since infinite mixtures of Gaussians can approximate any distribution. 
Notice that $\bm \Xi_t$ varies across components in the DPM. Depending on the estimated values for $\delta_t$ this allows for a wide variety of behavior (i.e. structural breaks, regime switching, outliers, etc.) in the contemporaneous relationships between the elements in $\bm y_t$. 

By additionally allowing for SV (via our specification for $\bm \Omega_t$) we have an error process of great flexibility allowing for impulse responses and other structural features to differ over time in a way that is estimated from the data. That is, both $\delta_t$ and $\bm \Omega_t$ allow for different types of parameter change, the latter only changing the error variances and being of a smooth nature, with the former additionally relating to the error covariances and allowing for more abrupt types of regime change or structural break. To see this feature, notice that a typical main diagonal element of $\bm \Xi_t$ is:
\begin{equation*}
   [\bm \Xi_t]_{ii} =  \sigma^2_{ii, \delta_t} + \omega_{it}.
\end{equation*}
In this equation, $\sigma^2_{ii, \delta_t}$, the $(i, i)^{th}$ element of $\bm \Sigma_{\delta_t}$, changes abruptly over time and is thus capable of handling large  outliers whereas $\omega_{it}$ changes smoothly and thus captures slowly varying trends in the error variances. Using only the latter implies that in the presence of large shocks, the SV model would only slowly adapt and would thus imply a higher variance when the large shock has already faded out. This model  resembles the SV with outliers model of \cite{carriero2022addressing}.

The conditional representation of the model also gives insights on how the DPM handles location shifts in the shocks. To see this, note that (\ref{BNPVAR}) allows for the intercept to change over time in a non-parametric manner. Models with time-varying intercepts are common in macroeconomics (see, e.g., \cite{StockWatson2007inflation} and \cite{Petrella}). Traditionally, intercepts have assumed to follow a random walk. However, our non-parametric treatment allows us to uncover the form of parameter change from the data \citep[see also][for a related but parametric treatment of time-varying parameter regressions]{hauzenberger2022fast}. 


It is also worth noting that an alternative model for non-parametric shocks would omit the additive error structure by setting $\bm v_t =0$ and simply have one vector of errors which is modelled using a DPM. But this apparently simpler form would both be more computationally demanding (i.e. since order invariant equation-by-equation estimation would be difficult to achieve) and would omit the SV process. That is, a DPM model for the errors on its own would be very flexible at modelling structural breaks and outliers, but the assumption that the DPM errors are independent over time means that is less able to model gradual changes in volatility. In contrast our model combines the benefits of a very flexible shock distribution with the gradual volatility change of an SV process.  


\section{Bayesian inference in the VAR with non-parametric shocks}\label{sec: bayes}
\subsection{Prior}
In this sub-section we describe our prior. We emphasize that the innovations in this paper relate to the parameters in the random effects. For the remaining parameters, any standard Bayesian priors can be used. In this paper, we use the Normal-Gamma prior of \cite{BrownGriffin} for the VAR coefficients \citep[see, e.g.,][]{huber2019adaptive} although any other common Bayesian VAR prior could be used (e.g. the Minnesota prior or a global local shrinkage prior such as the Horseshoe). We also adopt  standard priors for the parameters of the log-volatilities $\log \omega_{it}$. These are Gaussian priors on the unconditional mean $\mu_{\omega, j}$, (transformed) Beta priors on the persistence parameter $\rho_{\omega, j}$ and Gamma priors on the innovation variances $\sigma^2_{\omega, i}$, respectively.\footnote{More precisely, we set $\mu_{\omega,j} \sim \mathcal{N}(0, 10)$, $\frac{\rho_{\omega, j}+1}{2} \sim \mathcal{B}(25, 5)$ and $\sigma^2_{\omega, i} \sim \mathcal{G}(1/2, 1/2)$.} We also consider a homoskedastic version of our model where $\bm \Omega_t = \bm \Omega$ for all $t$ and in this case assume priors $\omega_{i} \sim \mathcal{G}^{-1}(a_\omega, b_\omega)$ for $i=1 \ldots M$ with $a_\omega = b_\omega = 10^{-3}$.

Our prior for the mean of the random effects is similar to one developed in \cite{M-W}. It is a Gaussian prior on $\bm \mu_j$ that shrinks the different elements of $\bm \mu_j$ towards a common location:
 \begin{equation*}
    \bm \mu_j \sim \mathcal{N}(\bm \mu_0, \bm B_0) \text{ for } j = 1, \dots, \infty,
\end{equation*}
with $\bm B_0 = \text{diag}(b_1, \dots, b_M)$ being a diagonal prior variance matrix with
\begin{equation*}
     b_j \sim \mathcal{G}(c_b, d_b).
\end{equation*}
The hyperparameters $c_b, d_b$ are greater than zero. This is the Normal-Gamma prior proposed in \cite{BrownGriffin}. If $c_b =1$ we obtain the LASSO \citep{park2008bayesian}. However, the LASSO is known to overshrink significant signals and undershrink irrelevant ones. Hence, we set $c_b=d_b=0.6$. This leads to a model that implies more shrinkage and flexible tail behavior. Since our prior is fully hierarchical, we also require another prior on $\bm \mu_0$. This is assumed to be $\mathcal{N}(\bm 0, c^{-1} \bm I_M)$ with $c \to 0$, yielding a non-informative prior. In all our empirical work, we set $c = 10^{-3}$ to render the prior relatively non-informative but proper.

The combination of a flexible shrinkage prior that forces the component-specific means towards a common location has implications on the clustering behavior of the mixture model. To illustrate this, let $\mu_{ji}$ denote the $j^{th}$ element of $\bm \mu_i$ for $i=k$ or $\tilde{k}$ (i.e. these are the two intercepts in the $j^{th}$ equation for two different components of the infinite mixture $k \neq \tilde{k}$). The prior above implies the following in terms of the distance between $\mu_{jk}$ and $\mu_{j \tilde{k}}$  \citep[see][]{YauHolmes}:
\begin{equation*}
        \frac{(\mu_{jk} - \mu_{j\tilde{k}})}{\sqrt{2}} \sim \mathcal{N}(0,  b_j).
\end{equation*}
Thus, our prior is centred over intercept homogeneity and $b_j$ controls the strength of this belief. If $b_j$ is close to zero, the intercepts collapse to a common value (which is $\mu_{0j}$, the $j^{th}$ element of $\bm \mu_{0}$). For larger $b_j$, we allow for more heterogeneity in the intercepts. This feature is crucial since the presence of the non-zero location parameter allows us to capture skewness in the shocks. If we use a non-informative prior on the component means we would risk overfitting the data. Our shrinkage prior effectively enables us to investigate how much asymmetries are in the data in a fully automatic manner.

For the covariance matrices for each component in the DPM, we use a conjugate Wishart prior on $\bm \Sigma^{-1}$ as this leads to order invariance within each component which implies order invariance in the VAR as a whole. Thus we assume 
\begin{equation*}
            \bm \Sigma^{-1}_k \sim \mathcal{W}(c_0 , \bm \Sigma_0^{-1}).
\end{equation*}
Note that we parameterize the Wishart such that the prior mean equals $c_0 \bm \Sigma_0^{-1}$ with $c_0$ being its degrees of freedom. The prior hyperparameters can be chosen in any way. In our empirical work, we use a relatively noninformative data-based prior inspired by the Minnesota prior.  In particular, we set the prior degrees of freedom as $c_0 = 2 (2.5 + (M-1)/2)$. The prior scaling matrix is estimated from the data and set as $\bm \Sigma_0 = \text{diag}(\hat{\sigma}^2_1, \dots, \hat{\sigma}^2_M)$ where $\hat{\sigma}^2_j$ are the OLS error variances obtained by running an AR($p$) model for $y_{jt}$. 
       
In terms of the weights in the DPM, we use a stick breaking process (SBP) prior on $\eta_j$. The SBP prior introduces additional auxiliary  random variables $\nu_j$ (called sticks) such that the weights are obtained sequentially:
\begin{equation*}
\eta_1 = \nu_1, \quad \eta_j = \nu_j \prod_{i=1}^{j-1} (1- \nu_j), \quad \nu_j \sim \mathcal{B}(1, \alpha).
\end{equation*}
The parameter $\alpha$ determines the clustering behavior of the mixture model. To see this, notice that the prior probability of forming a new cluster when assigning $\bm \varepsilon_t$ conditional on all $\bm \varepsilon_\tau (\tau \neq t)$ is \citep{lau2007bayesian, fruhwirth2019here}:
\begin{equation}
    \frac{\alpha}{T-1 + \alpha},
\end{equation}
and thus decreases in $T$. This implies that the DPM has the potential to create few larger clusters and then has a low probability of opening up new clusters that are populated by relatively few observations. In macroeconomic data, this behavior might be necessary to single out events that are are different from those produced by the DGP in normal times such as the Covid-19 pandemic. Since $\alpha$ crucially impacts this behavior, we treat it as an unknown parameter and estimate it from the data. We assume that $\alpha$ arises from  a Gamma distribution a priori, i.e., $\alpha \sim \mathcal{G}(2, 4)$ which implies a prior mean $0.5$ and a prior variance $0.125$. This choice was originally suggested by \cite{escobar1995bayesian} and encourages clustering behavior of the mixture model. 

\subsection{Posterior simulation of the VAR coefficients and random effects}
In this sub-section we describe how we simulate the VAR coefficients and random effects in more detail. The other steps are relatively standard and we sketch them in the next sub-section.  A key theme of this paper is computational efficiency and, to achieve this end, we need to draw the VAR coefficients one equation at a time. This requires knowledge of the random effects which serve to establish correlations across the shocks. Accordingly, we describe how we draw the VAR coefficients and the random effects in more detail.

Conditional on the random effects $\{\bm w_t\}_{t=1}^T$ and the estimates of $\{\bm Q_{\delta_t}\}_{t=1}^T$ we draw the equation-specific VAR coefficients $\bm A_{i}$ from
\begin{equation*}
        \bm A_{i}|\bullet \sim \mathcal{N}(\bm m_{i}, \bm V_{a, i}), \quad i=1,\dots, M,
\end{equation*}
where $\bullet$ denotes all arguments necessary to define the full conditional posterior distribution and:
\begin{align*}
        \bm V_{a, i} &= (\tilde{\bm X}'_i \tilde{\bm X}_i + \underline{\bm V}^{-1}_{a, i})^{-1},\\
        \bm m_i &= \bm V_{a, i} \tilde{\bm X}'_i \tilde{\bm Y}_i,
\end{align*}
denote the posterior moments with $\tilde{\bm X}_i$ being a $T \times K$ matrix with typical $t^{th}$ row $\bm X_t/\sqrt{\omega}_{it}$, $\underline{\bm V}_{a, i}$ denotes the prior variance matrix, $\tilde{\bm Y}_i$ denotes a $T \times 1$ vector with typical element given by $(y_{it} - \bm q'_{i \delta_t}\bm w_t)/\sqrt{\omega}_{it}$ and $\bm q_{i \bullet, \delta_t}$ denotes the $i^{th}$ row of $\bm Q_{\delta_t}$. For large models, the inversion of the posterior covariance (and computation of its Cholesky factor) becomes computationally cumbersome. However, in these cases (characterized by $T \ll K$) we can use the algorithm outlined in \cite{bhattacharya2016fast} that speeds up computation enormously. 

Next we describe the sampling steps involved in simulating the random effects in more detail. As stated in the previous sub-section, the model in (\ref{BNPVAR}) resembles a factor model with the particular structure on the factor loadings given in (\ref{efactor}).
We can use this representation to back out the conditional posterior distribution of $\bm w_t$, $p(\bm w_t|\bullet)$. The key point to notice is that the random effects are conditionally independent over time and hence $p(\bm w_1, \dots, \bm w_T|\bullet) = \prod_{t=1}^T p(\bm w_t|\bullet)$ with time $t$ posteriors given by:
\begin{equation}
    p(\bm w_t|\bullet) = \mathcal{N}(\overline{\bm w}_t, \overline{\bm V}_{w, t}), \label{post_w}
\end{equation}
and moments given by
\begin{align*}
    \overline{\bm V}_{w, t} &= (\tilde{\bm Q}_{\delta_t} \tilde{\bm Q}'_{\delta_t} + \bm I_M)^{-1},\\
    \overline{\bm w}_t &= \overline{\bm V}_{w, t} (\tilde{\bm Q'}_{\delta_t}\tilde{\bm \varepsilon_t}),
\end{align*}
where $\tilde{\bm Q}_{\delta_t} = \bm Q_{\delta_t} \bm \Omega_t^{-1/2}$ and $\tilde{\bm \varepsilon_t} =  \bm \Omega_t^{-1/2} \bm \varepsilon_t$.

\subsection{Full conditional MCMC sampling}
Our MCMC sampler is relatively straightforward and contains several steps which involve standard full conditional distributions.  Here, we summarize the different updating steps necessary to sample from the joint distribution of the latent states and coefficients of the model.

 Our sampler iterates between the following steps:
\begin{enumerate}
    \item Sample $\bm A_i|\bullet$ for each equation from $p(\bm A_i|\bullet) = \mathcal{N}(\bm m_i, \bm V_{a, i})$ as described in the previous sub-section.
    
    \item Sample $\bm \Sigma^{-1}_k$ from $p(\bm \Sigma_k|\bullet) = \mathcal{W}(\overline{c}_k, \overline{\bm \Sigma}_k)$. The posterior degrees of freedom are $\overline{c}_k = c_0 + T_k/2$ where $T_k = \sum_{t=1}^T \mathbb{I}(\delta_t = k)$ denotes the number of observations allocated to cluster $k$. The posterior scaling matrix is given by 
    \begin{equation*}
     \overline{\bm \Sigma}_k = \frac{1}{2}{\sum_{t: \delta_t=k}(\tilde{\bm y}_t - \bm \mu_{\delta_t}) (\tilde{\bm y}_t  - \bm \mu_{\delta_t})'} + \bm \Sigma^{-1}_0,
    \end{equation*}
    where $\tilde{\bm y}_t = \bm y_t - \bm A' \bm x_t - \bm v_t$.
    
    \item Sample $\bm \mu_k$ from $p(\bm \mu_k|\bullet) = \mathcal{N}(\overline{\bm \mu}_k, \overline{\bm V}_{\mu, k})$. The posterior moments of the random intercept terms are given by
    \begin{align*}
         \overline{\bm V}_{\mu, k} &= \left(\bm \Sigma_k^{-1} T_k +  \bm B_0^{-1}\right)^{-1}, \\
         \overline{\bm \mu}_k &= \overline{\bm V}_{\mu k} \left(\bm \Sigma_k^{-1} \sum_{t: \delta_t=k} \tilde{\bm y}_t + \bm B_0^{-1} \bm \mu_0\right).
    \end{align*}
    \item Sample $\bm \mu_0$ from $p(\bm \mu_0|\bullet) = \mathcal{N}(\overline{\bm \mu}_0, \overline{\bm V}_{\mu, 0})$ with
    \begin{align*}
        \overline{\bm V}_{\mu, 0} = \frac{1}{J} \bm B_0, \quad \overline{\bm \mu}_0 = \frac{1}{J}\sum_{j=1}^J \bm \mu_j.
    \end{align*}
    \item Sample $\bm w_t$ from $p(\bm w_1, \dots, \bm w_T| \bullet) = \prod_{t=1}^T \mathcal{N}(\overline{\bm w}_t, \overline{V}_{w t})$ with moments given below \autoref{post_w}.

    \item Sample $b_i$ (for $i=1, \dots, M$) from $p(b_j|\bullet) = \text{GIG}(p_0, p_K, z_j)$ with $p_0 = 2 d_b$, $p_K = c_b - J/2$ and $z_j = \sum_{j=1}^J (\mu_{j, i}  - \mu_{0, i})^2 $ and $\text{GIG}$ denoting the generalized inverse Gaussian distribution.\footnote{The density of the GIG is given by $f(x) = x^{\lambda -1} e^{-\frac{1}{2}(\chi/x + \psi x)}$.}
    
    \item Sample $\omega_{1t}, \dots, \omega_{Mt}$ from $p(\omega_{1t}, \dots, \omega_{Mt}|\bullet)$. For the SV case, $\omega_{j1}, \dots, \omega_{jT}$ (for $j=1,\dots, M$) and the parameters of the state equation are drawn using the algorithm outlined in \cite{kastner2014ancillarity} and implemented in the \texttt{R} package \texttt{stochvol}. In case we assume homoskedasticity, we simply draw $\omega_{j}$ from an inverse Gamma posterior.
    \item Sample $\nu_1, \dots, \nu_{J-1}$ from $p(\nu_1, \dots, \nu_{J-1}|\bullet) = \prod_{j=1}^J \mathcal{B}\left(1 + T_j, \alpha + \sum_{l=j+1}^J T_l\right)$.
    \item Sample $\delta_1, \dots, \delta_T$ from $p(\delta_1, \dots, \delta_T|\bullet) = \prod_{t=1}^T p(\delta_t|\bullet)$  using the Slice sampler \citep{kalli2011slice} in two steps. First, let $u_t|\delta_t \sim \mathcal{U}(0, \zeta_{\delta_t})$ denote a set of auxiliary random variables with $\zeta_k = (1-w) w^{k-1}$ and $w=0.8$. Then, conditional on $u_t$, we simulate $\delta_t$ from its discrete distribution as follows: 
    \begin{equation*}
        \text{Prob}(\delta_t = k | \bullet) \propto \frac{\mathbb{I}(u_t < \xi_k)}{\xi_k} ~\eta_k ~f_\mathcal{N}(\tilde{\bm y}_t |\bm \mu_k, \bm \Sigma_k).
    \end{equation*}
    \item Sample $\alpha$ using a Metropolis-Hastings step. 
\end{enumerate}
Notice that $J$ is a truncation parameter that determines the effective number of regimes which is obtained by solving $1-\sum_{j=1}^J \zeta_j < \text{min}(u_1, \dots, u_T)$. This implies that our infinite mixture model becomes effectively finite dimensional and thus computationally tractable. 

In most our applications we repeat this algorithm 20,000 times and discard the first 10,000 draws as burn-in. For the actual data application and in simulations we did not encounter any mixing issues if we consider functions of the parameters such as impulse responses or forecast densities.

Before applying our model to the data, it is worth stressing that the mixture model is not identified with respect to relabeling the latent discrete indicators. This does not cause any issues if interest is exclusively on either unconditional impulse responses or forecast distributions. In our structural application, we discuss impulse responses conditional on a given cluster. To avoid label switching in this case, we introduce restrictions that ensure a unique ordering ex-post. More details are provided below.

\subsection{Computational aspects of our algorithm}\label{sec: computational}
The algorithm described in the previous sub-section is efficient and has the same complexity as the original (but incorrect) version of the algorithm proposed in \cite{CCM2019} and its corrected variant proposed in \cite{carriero2022corrigendum}.\footnote{Notice that \cite{CCM2019} do not use the efficient sampling algorithm of \cite{bhattacharya2016fast} to simulate from the equation-specific coefficient posteriors if $T < K$. } This is because we sample from the equation-specific posteriors. This implies that the posterior covariance matrices are $K \times K$ and inversion of such matrices has computational complexity $O(M^4)$. A similar computational advantage can be obtained by applying the algorithm outlined in \cite{KastnerHuber} which uses a factor model to render equation-by-equation estimation possible. In any case, all these algorithms are much more efficient than the one based on treating the VAR as a full system of equations in large dimensions. 

\begin{figure}
    \centering
    \includegraphics[scale=0.55]{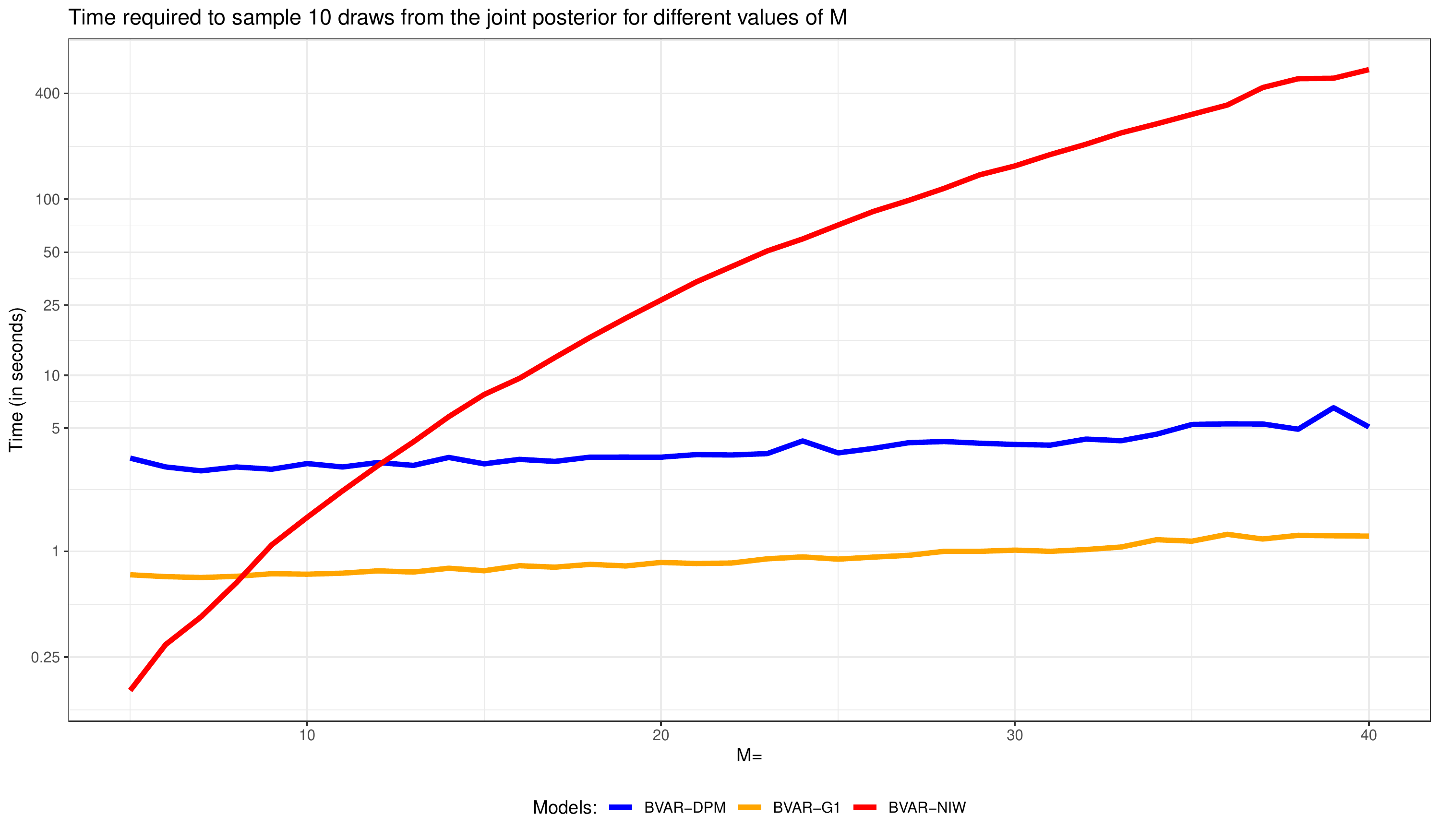}
    \caption{Comparison of computation times between the BVAR-DPM, BVAR-G1 and the BVAR with an independent Normal-Wishart prior (BVAR-NIW) for $M \in \{5, \dots, 40\}$. The y-axis is in log scale. }
    \label{fig:comps}
\end{figure}

 Figure \ref{fig:comps} illustrates the magnitude of efficiency improvements of equation-by-equation estimation relative to full system estimation and the increase in computation required to add the DPM. It compares our model (BVAR-DPM) to a restricted version of it which is Gaussian and obtained by setting $G=1$ (BVAR-G1) and a BVAR estimated under a non-conjugate independent inverse Wishart prior (BVAR-NIW). All models feature a single lag. Figure \ref{fig:comps} compares the computation times necessary to generate $10$ draws from the posterior. Note that the figure is in log-scales.

The figure shows that BVAR-DPM and BVAR-G1 are much faster than full-system estimation when $M$ exceeds $8$ (in the case of BVAR-G1) and when $M$ exceeds $12$ (in the case of the BVAR-DPM). In larger panels, the BVAR-NIW becomes much slower and the computational burden increases substantially. Differences between the blue and orange line reflect the additional computational burden from adding the DPM piece to the model. In most empirical work, the effective number of regimes is small (i.e. the infinite mixture reduces to a four-components mixture of Gaussians). This implies that if the true number of regimes is small, the computational burden would also decrease under the DPM specification. 

The algorithms of \cite{CCM2019} and \cite{carriero2022corrigendum} have the same computational complexity as our approach that sets $G=1$. In this case, computation times would be very similar (and the shape of the curves would, in fact, be identical) and our algorithm thus scales as well as theirs. However, it is worth stressing that ours is order invariant, can have non-parametric errors, and is applicable to any conditional mean function.


\section{Artificial data exercise}\label{sec: simulation}

We illustrate the merits of our approach by simulating data from a set of different DGPs. These DGPs differ both in terms of model size and the error distributions. With respect to model size, we consider three  sizes that capture typical situations in applied macroeconomic work. The smallest DGP has $M=5$ endogenous variables, the medium-sized DGP features $M=10$ and the largest includes $M=20$ endogenous variables. 

With respect to the error distributions we consider four different shock assumptions. Three of these four feature substantial departures from homoskedasticity and normality while one assumes Gaussian and homoskedastic shocks. All of these DGPs assume that the conditional mean of the process is given by:
\begin{equation*}
    \bm y_t = \bm A \bm y_{t-1} + \bm \epsilon_t, \quad \bm y_0 = \bm 0, \quad t=1,\dots, 250,
\end{equation*}
where $\bm A$ has diagonal elements $A_{ii}=0.75$ and off-diagonal elements sampled from a Gaussian distribution $A_{ij} \sim \mathcal{N}(0, 0.1^2) (i\neq j)$. To ensure stationarity we reject draws of $\bm A$ which imply an unstable model.

The errors  $\bm \epsilon_t$ differ across the DGPs in the following four ways.
    \begin{enumerate}
        \item The first DGP assumes that the shocks follow a multivariate skew Gaussian distribution: $\bm \epsilon_t \sim s\mathcal{N}(\bm \mu_t, \bm W, \bm \kappa)$ where $\bm \mu_t$ is a location parameter that equals $\bm 0$ with probability $98.5$ percent and $-3 \times z$, $z \sim \mathcal{U}(0, 1)$ with probability $1.5$ percent of the time. This increases the complexity of this error distribution further by capturing a situation where we have a few, large and negative shocks. The matrix $\bm W$ is created as follows. We specify a lower uni-triangular matrix $\bm U$ with $u_{ij} \sim \mathcal{N}(0, 0.1^2)$ for $i=2, \dots, M; j =1, \dots, M-1$ and then set $\bm W= \bm U \bm U'$. The parameter $\bm \kappa$ is obtained by simulating from a zero-mean Gaussian distribution with variance $9$ and then rounded to the nearest integer. 
    
        \item For the second DGP we assume that $\bm \varepsilon_t$ is multivariate Student t with three degrees of freedom and covariance matrix $\bm W$.
        
        \item The third DGP has a common stochastic volatility specification \citep{carriero2016common}. We assume that the shocks are Gaussian distributed with zero mean and time-varying covariance matrix $\bm W_t = e^{\mathfrak{s}_t} \times \bm W$. The log-volatility process $\mathfrak{s}_t$ evolves according to a random walk with innovation variance $0.25^2$
        \item Finally, we also consider a homoskedastic DGP that assumes that $\bm \varepsilon_t \sim \mathcal{N}(\bm 0, \bm W)$.
    \end{enumerate}
For comparison, we consider the BVAR-DPM and the BVAR-DPM with $G=1$ and set the number of lags equal to 5.   All models are estimated with homoskedastic and SV measurement errors. This combination allows us to analyze whether the BVAR-DPM is capable of recovering non-Gaussian features in the DGP without overfitting. In particular, the final DGP can be used to focus on the question whether adding the non-parametric component to the model leads to a deterioration in estimation accuracy or whether the DPM can recover the simple Gaussian case without overfitting.
\begin{table}[ht!]
\centering
\begin{threeparttable}
\caption{Simulation results}
\begin{tabular}[t]{lrrrrrrrr}
\toprule
\multicolumn{1}{c}{} & \multicolumn{2}{c}{Skew} & \multicolumn{2}{c}{ t } & \multicolumn{2}{c}{SV} & \multicolumn{2}{c}{Homosk.} \\
\cmidrule(l{3pt}r{3pt}){2-3} \cmidrule(l{3pt}r{3pt}){4-5} \cmidrule(l{3pt}r{3pt}){6-7} \cmidrule(l{3pt}r{3pt}){8-9}
$M \downarrow$ & DPM & $G=1$ & DPM & $G=1$ & DPM & $G=1$ & DPM & $G=1$\\
\midrule
\addlinespace[0em]
\multicolumn{9}{l}{\textbf{Homosk.}}\\
5 & 1.49 & 1.54 & 1.19 & 1.93 & 4.96 & 5.71 & 1.58 & 1.58\\
10 & 1.52 & 1.56 & 1.22 & 1.71 & 4.83 & 5.16 & 1.65 & 1.65\\
20 & 1.45 & 1.48 & 1.84 & 1.92 & 3.81 & 3.90 & 1.56 & 1.55\\
\addlinespace[0em]
\multicolumn{9}{l}{\textbf{SV}}\\
5 & 1.86 & 1.99 & 1.36 & 1.57 & 1.93 & 1.80 & 1.94 & 1.92\\
10 & 1.78 & 1.80 & 1.23 & 1.35 & 1.71 & 1.68 & 1.82 & 1.81\\
20 & 1.59 & 1.60 & 1.21 & 1.22 & 1.57 & 1.56 & 1.67 & 1.67\\
\bottomrule
\end{tabular}
\caption*{\footnotesize \textbf{Notes}: This table shows  mean absolute errors (MAEs) between the posterior median of the VAR coefficients and the true VAR coefficients.  DPM is the BVAR with a DPM of Gaussians and $G=1$ is a BVAR with a set of random coefficients (G=1).}
\label{tab:MAE}
\end{threeparttable}
\end{table}

To rank models in terms of estimation accuracy we focus on the mean absolute error (MAE) between the true set of coefficients and the posterior median of the estimated VAR coefficients.  Table \ref{tab:MAE} shows the accuracy of the BVAR-DPM versus the BVAR with Gaussian errors in terms of their estimation of the VAR coefficients. All results are means across MAEs from 50 replications from each of the DGPs.

In general, our results indicate that if the Gaussian-errored model is mis-specified,  the DPM model is consistently more accurate. But even for cases where the BVAR $G=1$ is correctly specified (i.e. with the homoskedastic model and homoskedastic DGP and with the SV model with the SV DGP) our BVAR-DPM is only slightly less accurate. This illustrates how the BVAR-DPM can successfully model features such as skewness and fat tails, but when the DGP does not have such features it can successfully uncover the underlying Gaussian model with very little over-fitting.  When we consider differences across model sizes, we find that accuracy gains from setting $G= \infty$ vis-\'{a}-vis $G=1$ decline with larger information sets. This finding points towards the fact that large models can soak up non-Gaussian features in the data. 

\begin{table}[t!]
\centering
\begin{threeparttable}
\caption{Effective number of clusters}
\begin{tabular}[t]{rcccc}
\toprule
\multicolumn{1}{c}{} & \multicolumn{1}{c}{Skew} & \multicolumn{1}{c}{t} & \multicolumn{1}{c}{SV} & \multicolumn{1}{c}{Homosk.} \\
\midrule
\addlinespace[0em]
\multicolumn{5}{l}{\textbf{Homosk.}}\\
5 & 2.03 & 3.98 & 2.22 & 1.00\\
10 & 2.05 & 2.51 & 1.92 & 1.00\\
20 & 2.09 & 1.00 & 1.42 & 1.01\\
\addlinespace[0em]
\multicolumn{5}{l}{\textbf{SV}}\\
5 & 2.10 & 4.22 & 2.65 & 1.00\\
10 & 2.08 & 2.71 & 1.54 & 1.00\\
20 & 2.10 & 1.15 & 1.64 & 1.03\\
\bottomrule
\end{tabular}
\caption*{\footnotesize \textbf{Notes:} This table shows the effective number of clusters. Numbers represent means over the different replications from the DGP.}
\label{tab:num_comps}
\end{threeparttable}
\end{table}

When we use the DPM we can infer the effective number of clusters. This is achieved as follows. For the $i^{th}$ run of our MCMC algorithm, we compute
\begin{equation*}
    G^{(i)} = \sum_j^J \mathbb{I}\left(T^{(i)}_j > 0\right).
\end{equation*}
We can then compute the posterior median of these runs to obtain an estimate of the number of clusters. Table \ref{tab:num_comps} shows the mean over these posterior median estimates across the different realizations from the DGP.  Note that, when the DGP is Gaussian and the DPM extension is unnecessary, our algorithm is correctly selecting $G=1$. When the DGP is skewed, the DPM is modelling this with a mixture of $G=2$ distributions, regardless of VAR dimension or whether the DGP has SV or not. When the DGP involves a Student t distribution (and to a lesser extent when the DGP is SV) we are finding an interesting pattern where the number of clusters is inversely related to the VAR dimension. This is consistent with our conjecture that, as the VAR dimension increases and more explanatory variables appear on the right hand side of each equation, the extra variables can fit some of the fat tailed behavior of the DGP. This tradeoff between VAR dimension and the need for a non-parametric distribution is something we find in our forecasting exercise and we will say more about this issue then. 

\section{Empirical application using US data}
\subsection{Data and specifications}
We use US quarterly macroeconomic data from 1960Q1 to 2022Q1 taken from the FRED database, see \cite{mccracken2020fred}. A full list of variables along the transformations we use is given in the Data Appendix. In our forecasting exercise, we evaluate forecast performance beginning in $1977Q1$ and rely on the iterated method of forecasting and consider forecast horizons of one-quarter  and one-year. 

We work with small ($M=4$), medium ($M=7$) and large ($M=28$) dimensional VARs with the variables included in each being given in the Data Appendix. All variables are transformed to be approximately stationary (with detailed information provided in the Data Appendix). In the forecasting exercise we evaluate the performance of the models by focusing on the variable-specific performance for GDP growth, the unemployment rate and inflation. These three variables form our set of focus variables. We choose a long lag length, $p=5$, and trust our shrinkage prior to prevent overfitting. 

In Sub-section \ref{sec: forecasting} we carry out a forecasting exercise to assess whether adding the DPM to the VAR improves predictive accuracy. To this end, we compare the performance of the model with non-parametric shocks (BVAR-DPM) to models with Gaussian errors which we obtain by taking our BVAR-DPM and setting $G=1$. This is the model described in Sub-section \ref{sec: linear}. All other specification details, including prior choice, are the same in all of our models. In this way, we can focus on the specific issue of what the use of the DPM adds to the model. Note that we are not also including a BVAR with full error covariance since, as established previously, it is computationally much more burdensome in larger models and is expected to give results very similar to the BVAR with $G=1$. We consider two versions of every model, one with SV and one homoskedastic. 

\subsection{Full sample analysis}
In order to illustrate the properties of our models, we begin by carrying out full sample estimation using a single model: the BVAR-DPM without SV using the medium data set.  This model works well empirically and allows for a straightforward comparison between impulse responses in the next sub-section.

An advantage of the DPM is that it can be used to estimate the effective number of components in the Gaussian mixture and our simulation results show that it does so accurately.
\begin{table}[h!]
\centering
\begin{threeparttable}
\caption{Probability of a given effective number of clusters}\label{tab: num_components}
\begin{tabular}{rrrrrrrrrr}
  \toprule
$G = $ & 1 & 2 & 3 & 4 & 5 & 6 & 7 & 8 & 9 \\ 
  \hline
  & 0.00 & 0.00 & 0.00 & 0.95 & 0.05 & 0.00 & 0.00 & 0.00 & 0.00 \\
   \bottomrule
\end{tabular}
\end{threeparttable}
\end{table}

Table \ref{tab: num_components} presents evidence relating to this. It provides quantitative information on the posterior distribution of the effective number of clusters. The table suggests that the BVAR-DPM-SV is allocating almost all of the probability to $G=4$ with the remaining probability allocated to $G=5$. Thus, we are finding no evidence in favour of the conventional Gaussian VAR with SV which has $G=1$.

\begin{figure}[t!]
    \centering
    \includegraphics[scale=0.45]{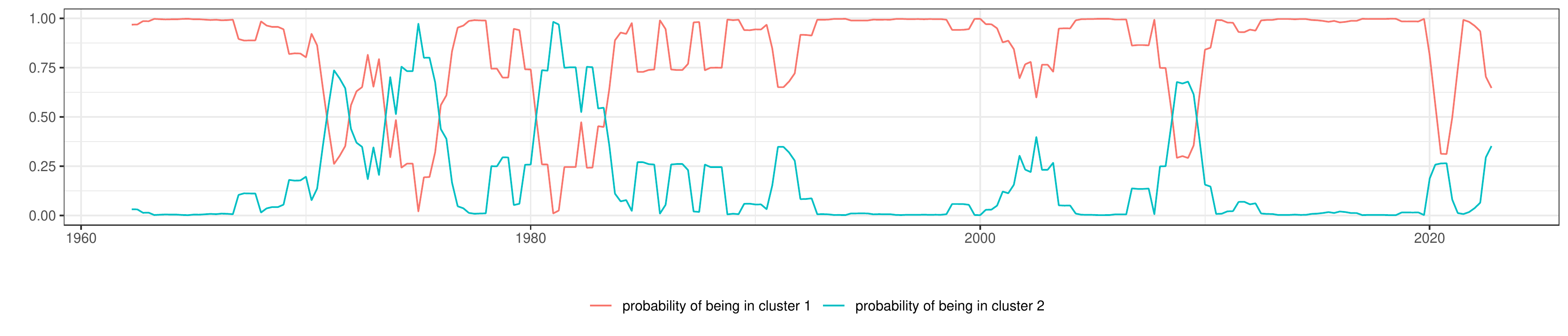}
    \includegraphics[scale=0.45]{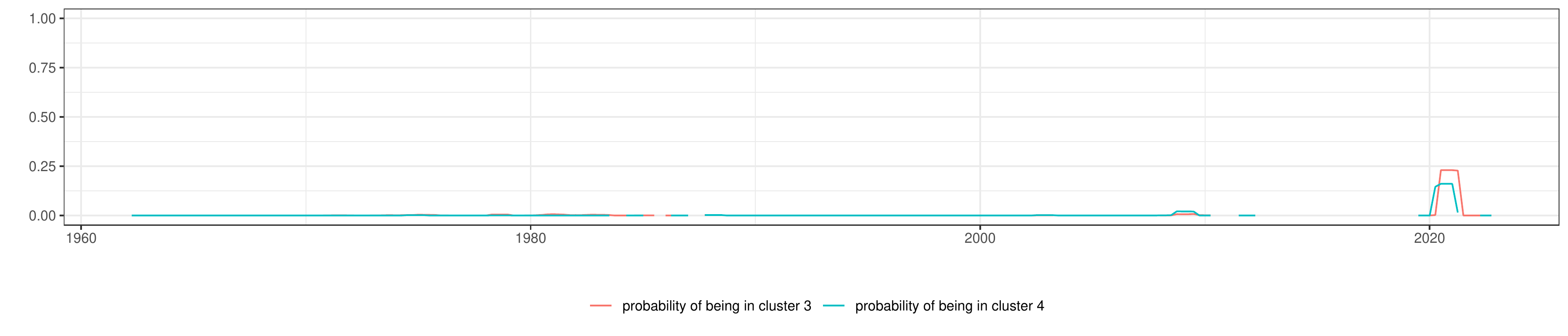}
    \caption{Probability of being in a given cluster over time}\label{fig: prob_over_time}
\end{figure}

The characteristics of the four clusters estimated by the DPM can be seen in Figure \ref{fig: prob_over_time}. This figure shows the posterior probability that $\delta_t = j$ for $j=1, \dots, 4$. To introduce persistence and simplify the discussion, we report yearly rolling averages of these probabilities.

Cluster 1 is predominant and holds with high probability in most periods. Cluster 2 is associated with times of high volatility such as periods in the early 1970s and around 1980 as well as the financial crisis and, to a lesser extent, the pandemic. Clusters 3 and 4 are completely associated with the pandemic. Another interesting point to note is that, much of the time Cluster 1 applies with probability near one. But in more volatile times like the financial crisis and the pandemic, no single cluster holds with probability near one. For instance, early in the Covid-19 pandemic each of the four clusters receives roughly equal probability. Thus, in normal times a single Gaussian distribution suffices to model the error distribution, but in less stable times a mixture of two or more Gaussians is required.

\begin{figure}[t!]
    \centering
    \includegraphics[scale=0.45]{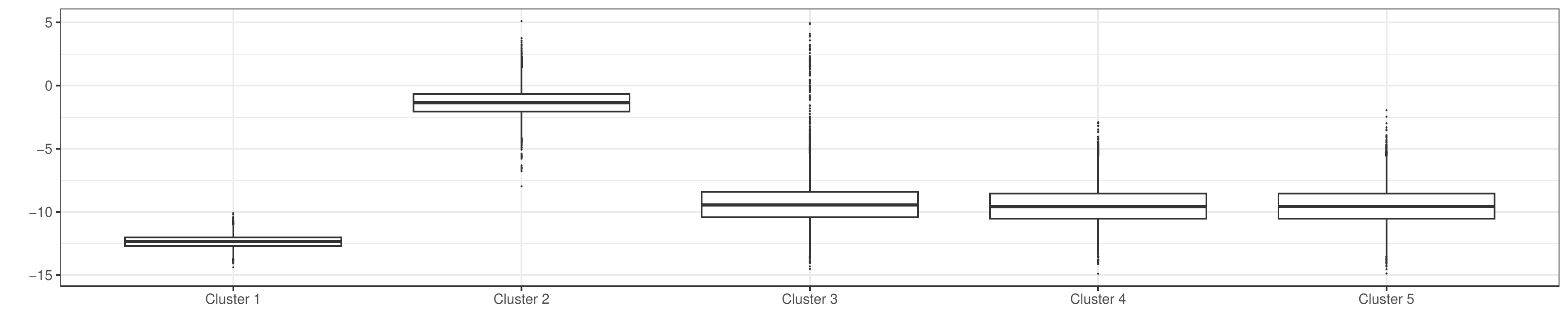}
    \caption{Log-Determinant of $\bm \Sigma_k$}
    \label{fig:log_det}
\end{figure}

A deeper understanding of the properties of the clusters can be obtained by looking at the posteriors of $\bm \Sigma_k$ for $k=1 \ldots 5$. Figure \ref{fig:log_det} contains box and whisker plots of the posterior of the log of the determinant of each $\bm \Sigma_k$. It can be seen that Cluster 1 is the low volatility cluster whereas Cluster 2 has much higher volatility (as evidenced by much larger log-determinants). Clusters 2 through 5 also have much more uncertainty (e.g. wider credible intervals) than Cluster 1.  

\subsection{Structural impulse responses to a monetary policy shock}
Next we investigate the dynamic effects of monetary policy shocks using a standard identification scheme with a Cholesky ordering where the Federal Funds rate is ordered above financial market variables and below real quantities. This ordering implies that real quantities belong to the 'slow-moving' block whereas financial markets are 'fast-moving' \citep{bernanke2005measuring}. To economize on space, we focus on the IRFs of the three focus variables (output growth, unemployment and  inflation).

A feature of our model is that, conditional on the mixture indicators $\delta_t$, our model can be interpreted as a constant parameter VAR with a fully time-varying covariance matrix. This implies that the structural form of the model features time-varying parameters. To see this, multiply (\ref{BNPVAR}) by $\bm \Psi_t^{-1}$, the inverse of the lower Cholesky factor of $\bm \Xi_t$, from the left. This yields:
\begin{equation*}
   \bm \Psi_t^{-1} \bm y_t =  \tilde{\bm A}_t \bm X_t + \tilde{\bm \varepsilon}_t.
\end{equation*}
Here we let $\tilde{\bm A}_t = \bm \Psi_t^{-1} \bm A$ and $\tilde{\bm \varepsilon}_t \sim \mathcal{N}(\bm 0_M, \bm I_M)$. The key implication is that the structural coefficients of the model are time-varying and all parameters in the structural form of the model change if the regime shifts.

\begin{figure}[t!]
    \centering
    \begin{subfigure}{1\textwidth}
            \caption{GDPC1}
        \includegraphics[scale=0.35]{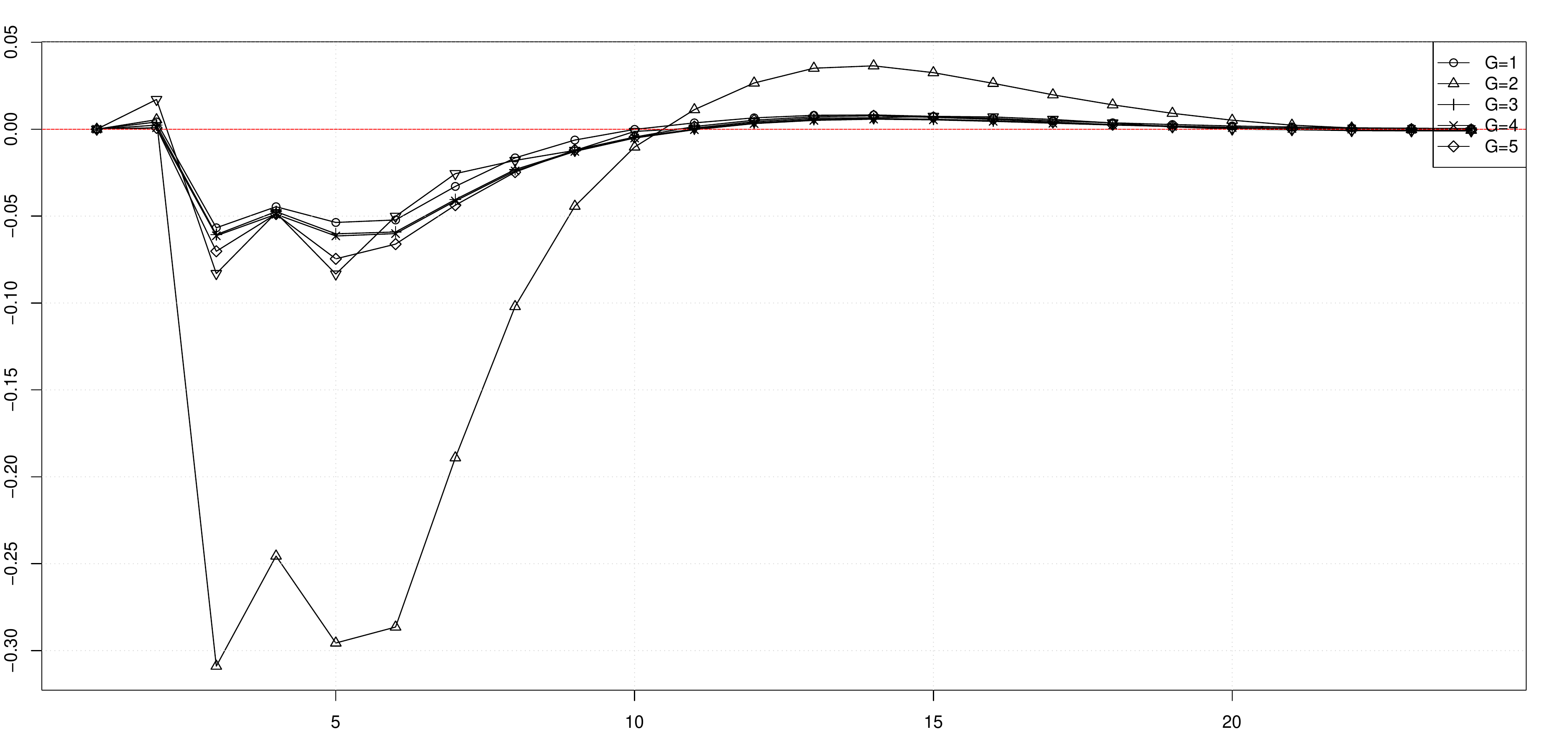}
        \label{fig:gdpc1}
    \end{subfigure}
        \begin{subfigure}{1\textwidth}
            \caption{UNRATE}
        \includegraphics[scale=0.35]{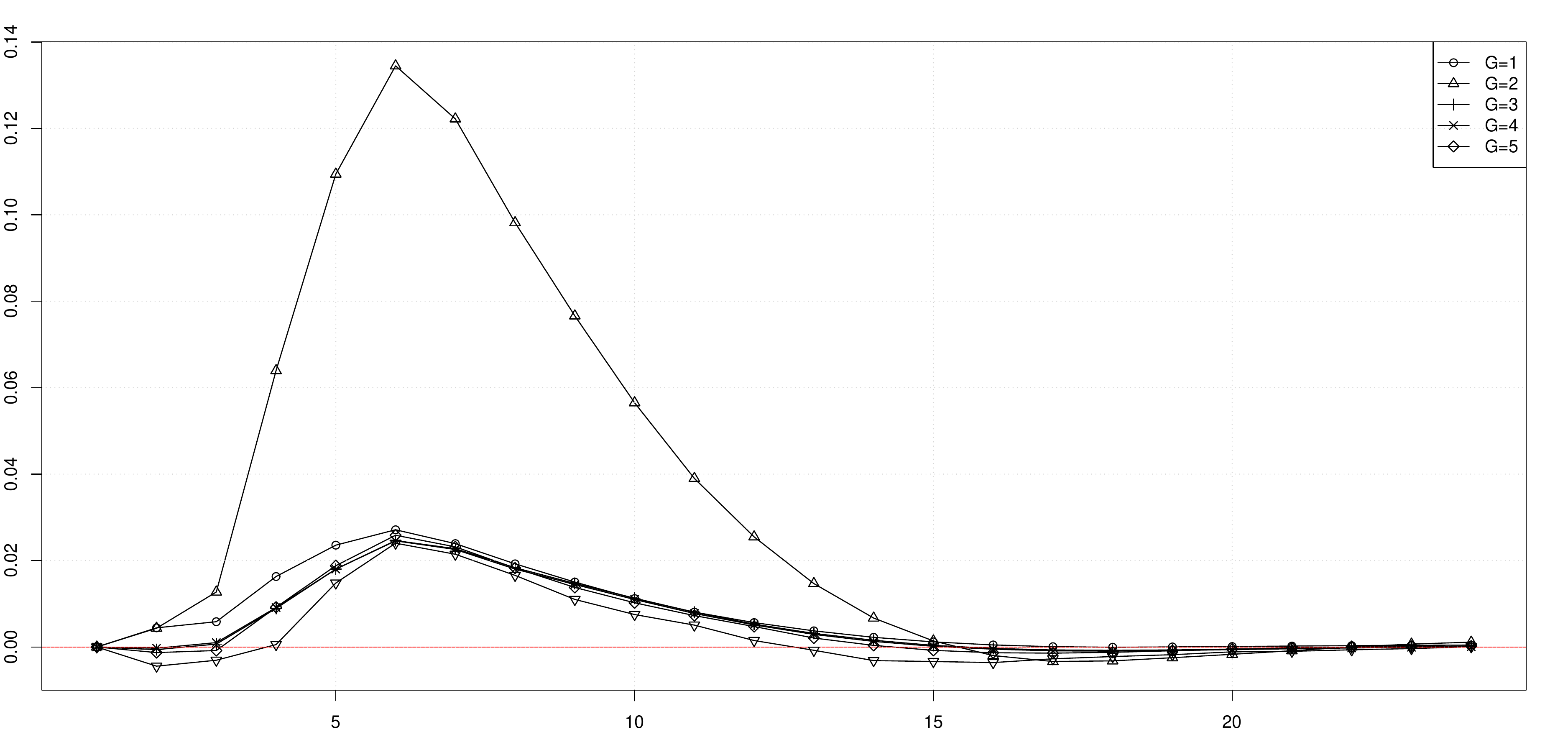}
        \label{fig:unrate}
    \end{subfigure}\\
        \begin{subfigure}{1\textwidth}
                \caption{CPIAUCSL}
        \includegraphics[scale=0.35]{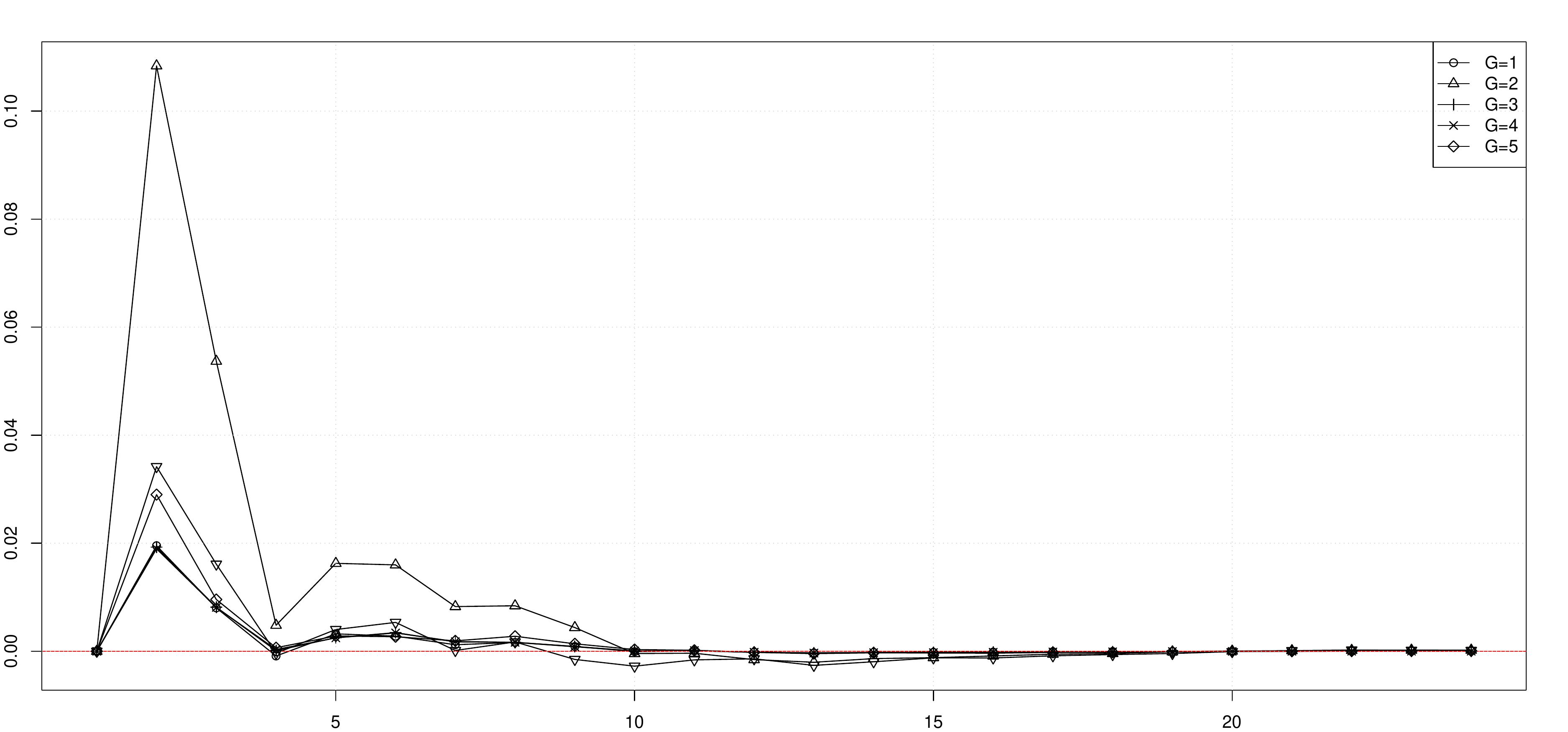}
        \label{fig:cpi}
    \end{subfigure}
  \caption*{\footnotesize \textbf{Notes}: The black lines show the posterior median of the impulse responses for the different clusters.}
        \caption{Median impulse responses to a monetary policy shock for the different clusters.}\label{fig: IRFS_median_regimes}
\end{figure}

We start our analysis by considering  median impulse responses over time. Since within each regime we have a standard VAR with constant coefficients, we compute the IRFs per regime. This gives a set of $G$ dynamic responses to a monetary policy shock. To analyze differences in shapes and magnitudes we focus on the posterior median in the main body of the paper. Results that include posterior credible intervals and the reactions of the other variables in $\bm y_t$  are provided in the Empirical Appendix. 

Figure \ref{fig: IRFS_median_regimes} presents posterior medians of the impulse responses implied by each cluster. In general, the model produces impulse responses that are consistent with our economic intuition. In response to unexpected increases in the policy rate, output growth declines and the unemployment rate increases. Inflation, unexpectedly, increases, pointing towards a price puzzle. It is worth stressing that this price puzzle is most pronounced in the second cluster, which is the high volatility cluster.

When we compare differences across clusters,  it can be seen that Cluster 2 stands out as implying very different impulse responses but in ways that are different for the different variables. For instance, all of the clusters are very similar for long-run impulse responses (e.g. greater than 10 quarters), Cluster 2 differs greatly from other clusters at short horizons (e.g. less than one year) for most of the variables. In particular, short-run responses in Cluster 2 appear to be much  more pronounced. This is driven by the fact that the variances of the structural shocks are much larger and the monetary shock implies a stronger impact reaction of the Federal funds rate (see panel (d) of Figure B.1 in the Empirical Appendix).  For medium-run responses, the second cluster also yields responses of GDP growth and unemployment rates which differ from the remaining clusters. For GDP growth, our results indicate an overshoot in real activity after around two years whereas the unemployment reaction appears to be much more persistent. This points towards differences in the transmission of monetary policy to  real activity in turbulent periods (i.e. periods that are allocated to Cluster 2).

\begin{figure}[t!]
    \centering
    \begin{subfigure}{.49\textwidth}
            \caption{GDPC1}
        \includegraphics[width=1\textwidth]{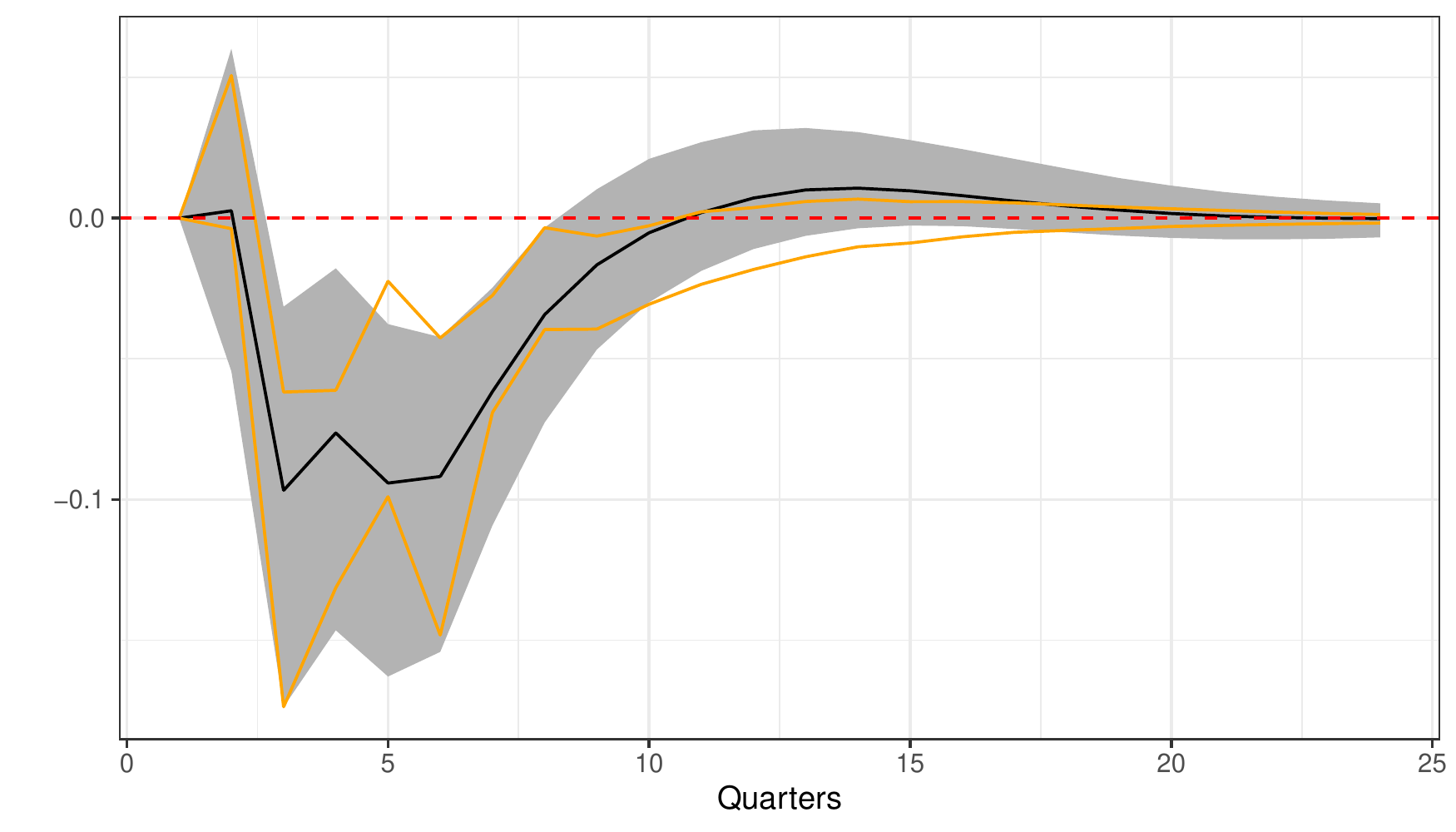}
        \label{fig:gdpc1}
    \end{subfigure}
        \begin{subfigure}{.49\textwidth}
                \caption{CPIAUCSL}
        \includegraphics[width=\textwidth]{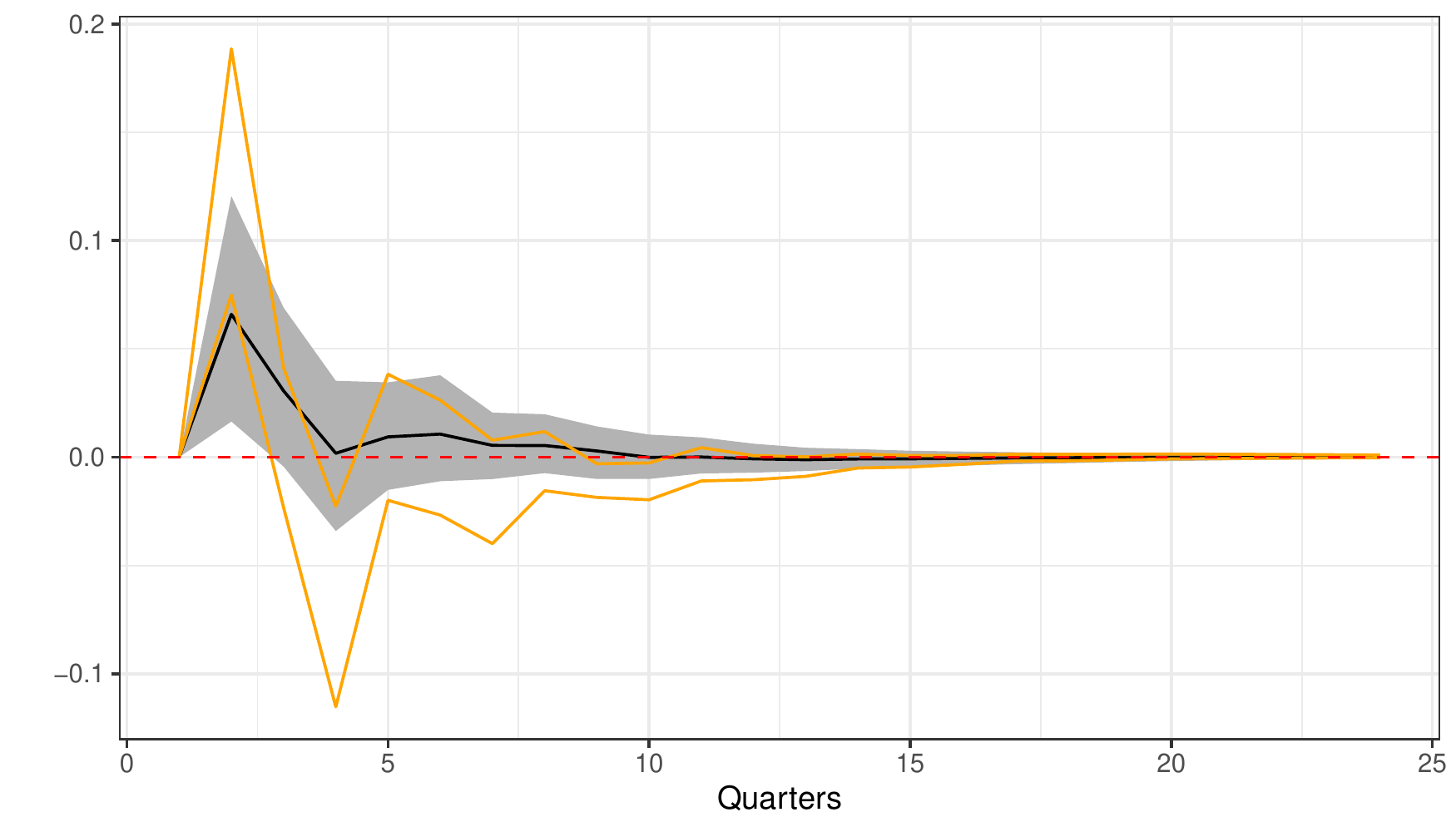}
        \label{fig:cpi}
    \end{subfigure}
    \begin{subfigure}{.49\textwidth}
            \caption{UNRATE}
        \includegraphics[width=\textwidth]{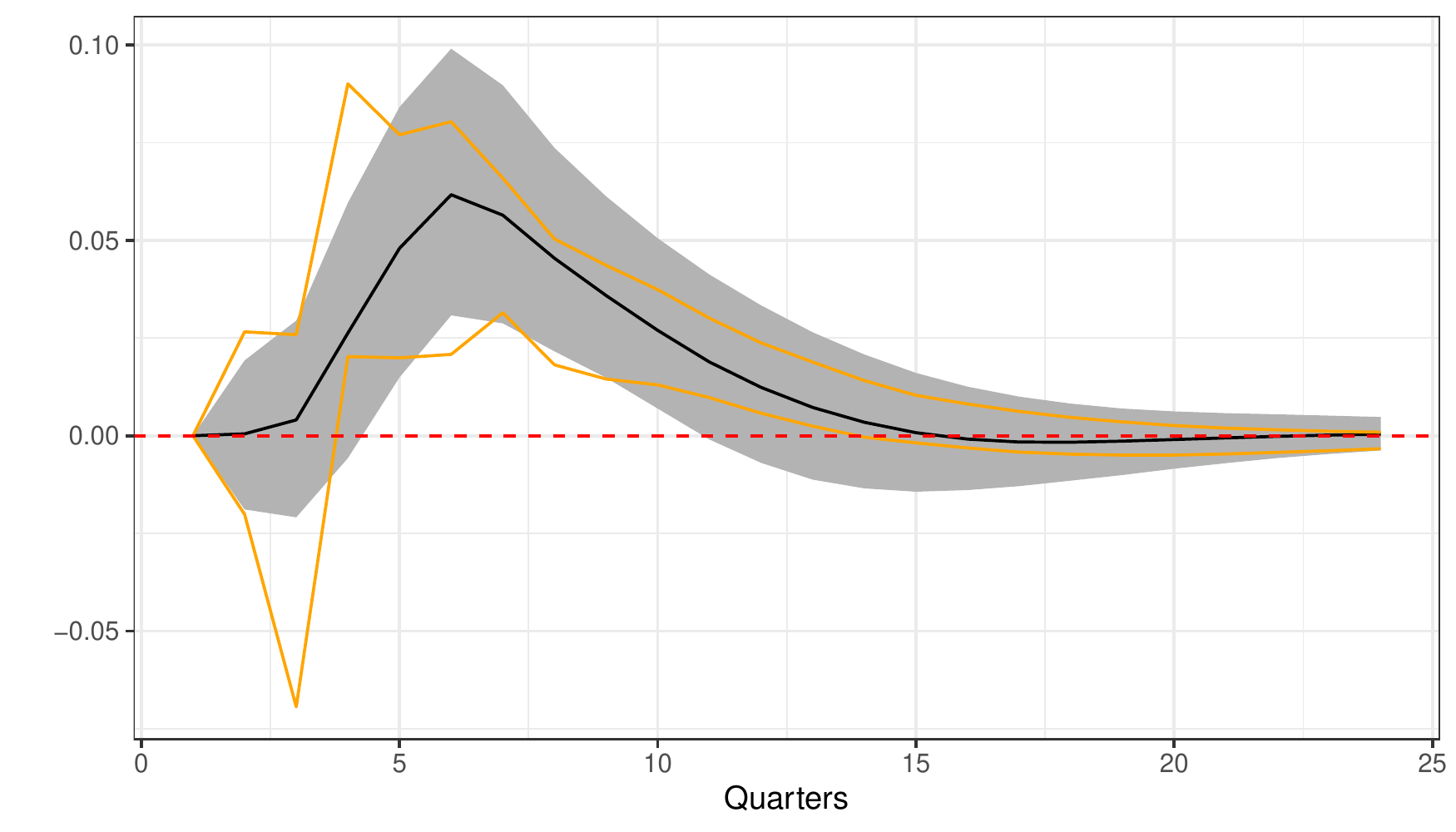}
        \label{fig:unrate}
    \end{subfigure}
    \caption*{\footnotesize \textbf{Notes}: The black lines show the posterior median of the impulse responses for the BVAR-DPM. The shaded region is a credible interval. The orange lines denote credible intervals for the Gaussian BVAR. Both intervals cover $16^{th}$ to $84^{th}$ percentiles.}
    \caption{Median impulse responses to a monetary policy shock, average across clusters.}\label{fig: overall_IRFs}
\end{figure}

Next we turn to the question about the overall effects of monetary shocks on the three focus variables. The overall impulse responses (i.e. averaged over the different clusters) for our core variables (i.e. the ones common to all VARs) are given in Figure \ref{fig: overall_IRFs}. These are calculated by taking the impulse responses for cluster $i$ and weighting it by the posterior mean of $\eta_i$. This figure also contains credible intervals for both our BVAR-DPM and the BVAR with $G=1$ so that the reader can gauge whether the differences in impulse responses between our model and the Gaussian-errored equivalent are substantial in a statistical sense.

The figure suggests that differences between the model with $G= \infty$ and $G=1$ are not substantial in the sense that the credible intervals of the BVAR-DPM include the ones of the BVAR in almost all cases. The main exception is the reaction of inflation. Here, we observe a much stronger immediate increase (i.e. more evidence for a price puzzle) but also a slightly more pronounced one-year-ahead decline in inflation. It is also worth stressing that short-run unemployment reactions of the BVAR with $G=1$ suggest substantial posterior mass of the IRFs are located below zero, suggesting a decline in the unemployment rate to a monetary tightening. The BVAR-DPM allocates appreciably less posterior evidence of declines in unemployment rates.

\subsection{Forecasting performance}\label{sec: forecasting}

In this sub-section we present the results of our forecasting exercise.  Our measure of point forecast performance is the mean squared forecast error (MSE) and our measure of density forecast performance is the average of log predictive likelihoods (LPL). Tail forecast performance is measured using absolute quantile scores (QSs).  Results are reported relative to a benchmark model which is the large BVAR with SV and Gaussian errors.

\begin{figure}[h!]
      \begin{subfigure}{1\textwidth}
            \caption{One-quarter-ahead}
        \includegraphics[scale=0.45]{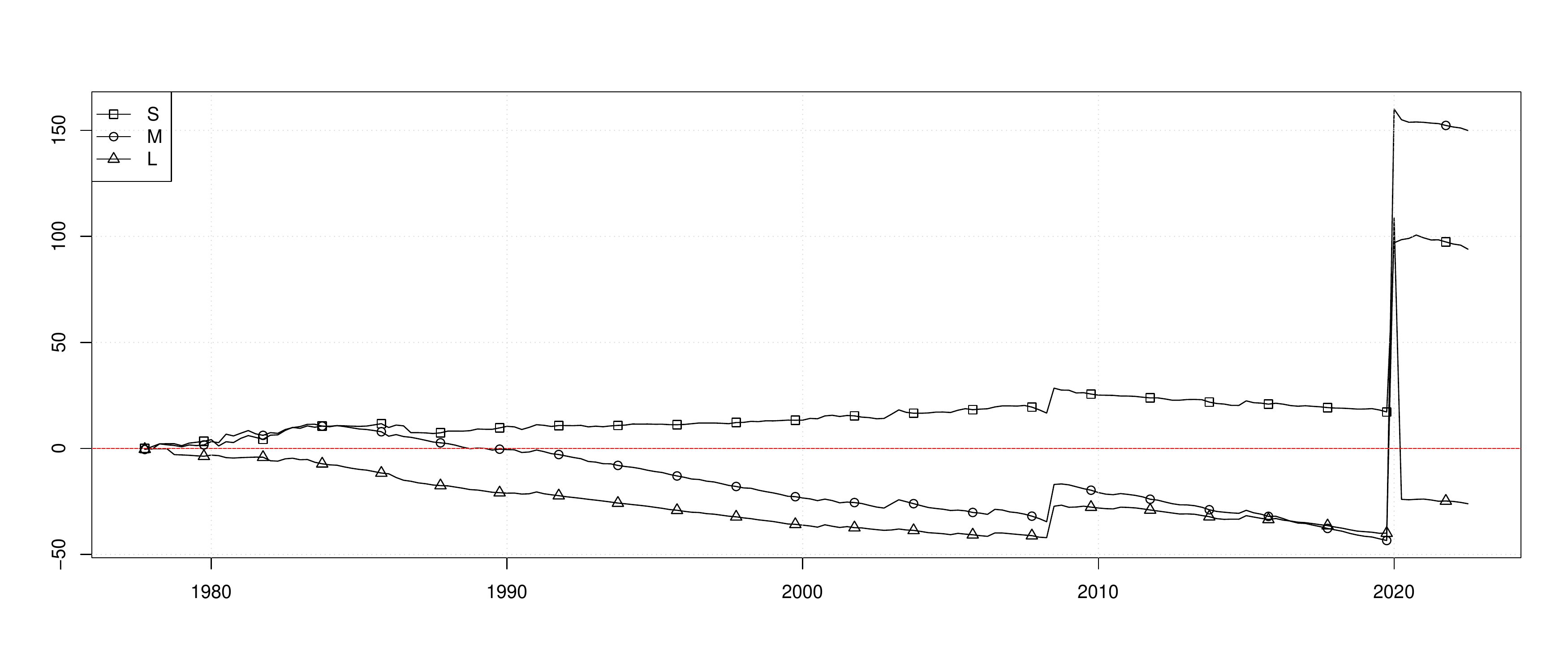}
        \label{fig:LPL_OT_1}
    \end{subfigure}
          \begin{subfigure}{1\textwidth}
            \caption{One-year-ahead}
        \includegraphics[scale=0.45]{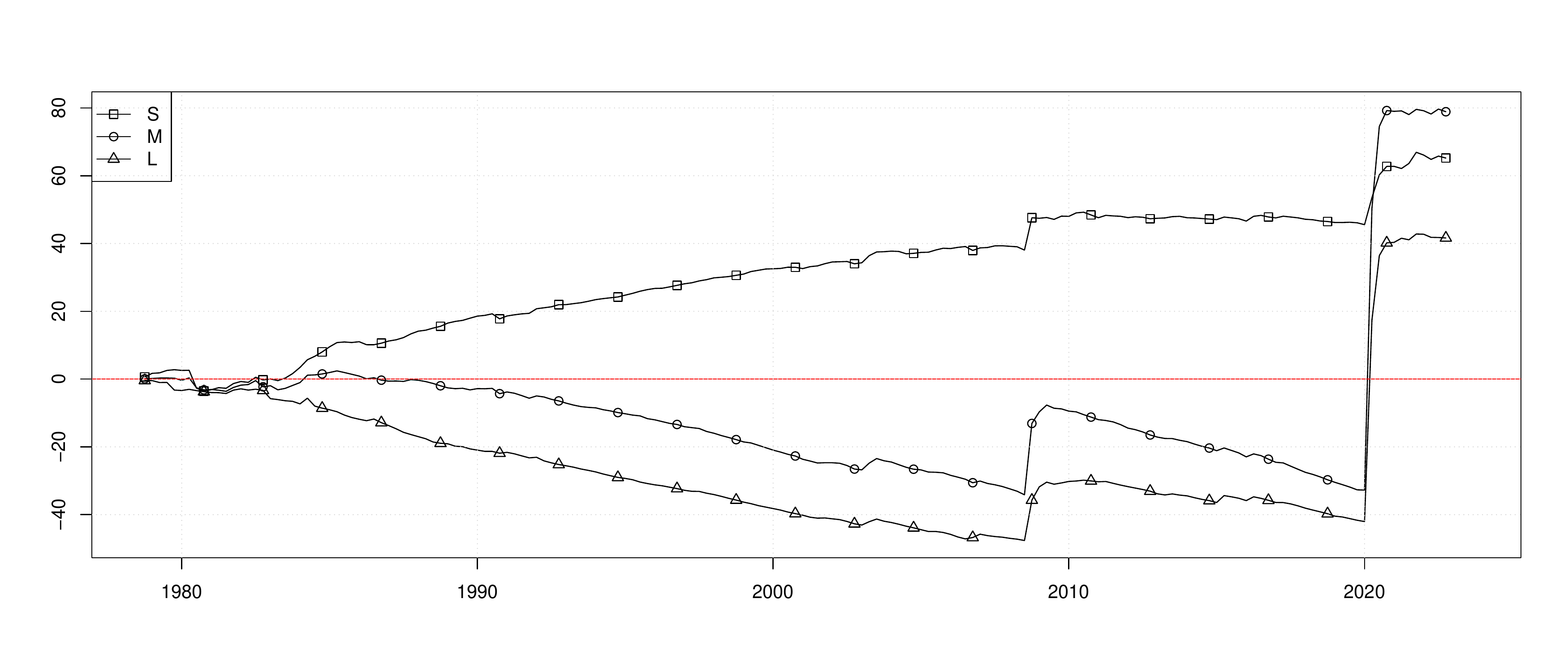}
        \label{fig:LPL_OT_4}
    \end{subfigure}
    \caption*{\footnotesize \textbf{Notes}: This figure shows cumulative log predictive likelihoods of the BVAR-DPM (across model sizes) relative to the large BVAR-G1 model. All models have SV.}
    \caption{Density forecasting performance over time}\label{fig: forecasts_over_time}
\end{figure}
Our discussion starts by considering the overall density forecasting performance for the three focus variables over the hold-out period (1977:Q1 to 2022:Q1).  Figure \ref{fig: forecasts_over_time} presents an overall summary of how our BVAR-DPM with SV models are performing over time. It contains the cumulative differences in the joint predictive likelihood for the three main variables being forecast (inflation, GDP growth and the unemployment rate) for the BVAR-DPM with SV and the BVAR-SV. Numbers greater than zero indicate that the DPM component of the model yields accuracy gains whereas negative numbers suggest that it hurts predictive performance.

When we consider the entire hold-out period, all of our BVAR-DPM with SV models beat the benchmark for both forecast horizons. In terms of VAR dimension, it is the medium VAR which forecasts best at both horizons. But this overall performance masks interesting patterns over time. It can be seen that the excellent performance of the DPM models is driven largely by the pandemic and the financial crisis. Clearly, adding the non-parametric treatment of shocks is particularly useful in such periods. This is because the likelihood of observing outliers under the predictive density increases if the training sample includes similar outliers. A model that does not appropriately capture outliers is not capable of producing abrupt changes in the predictive variance and thus underestimates the probability of tail events.

In tranquil periods,  the only DPM model which beats the benchmark is the small BVAR-DPM. Hence, we have a story where, in normal times, smaller, more parsimonious models suffice but in unusual times larger, non-Gaussian models forecast better. An advantage of the BVAR-DPM approach is that it can decide automatically when to introduce non-Gaussian errors and when to simply choose $G=1$ and use Gaussian errors.  

Next we zoom into variable-specific forecasting performance. We do so by focusing on MSE ratios and LPL differences between a given model and the large VAR with SV. Table \ref{tab: forecast_LPLMAE} present the relative MSEs and LPLs (in parentheses) for the three variables being forecast for both homoskedastic and heteroskedastic models, allowing for a variable-specific examination of forecast performance.  

\begin{table}[t!]
\centering
\caption{Forecasting performance: 1977:Q1 to 2022:Q1.}\label{tab: forecast_LPLMAE}
\begin{threeparttable}
\scalebox{0.98}{
\begin{tabular}[t]{llrrrrrr}
\toprule
& &  \multicolumn{2}{c}{GDPC1} & \multicolumn{2}{c}{UNRATE} & \multicolumn{2}{c}{CPIAUCSL} \\
 & & DPM & $G=1$ & DPM & $G=1$ & DPM & $G=1$\\
\midrule
\multicolumn{8}{l}{\textbf{One-quarter-ahead}}\\
Homosk. & S & 0.951 & 1.219 & 1.007 & 1.535 & 0.912 & 0.951\\
 &  & (-0.019) & (-0.055) & ( 0.550) & (-0.664) & ( 0.065) & ( 0.020)\\
 & M & 0.872 & 1.003 & 0.959 & 1.291 & 0.836 & 0.863\\
 &  & (-0.023) & (-0.014) & ( 0.965) & (-0.596) & ( 0.133) & ( 0.060)\\
 & L & 0.955 & 1.022 & 1.080 & 1.055 & 0.804 & 0.824\\
 &  & (-0.090) & (-0.165) & ( 0.073) & (-0.037) & ( 0.086) & ( 0.070)\\
\addlinespace
SV & S & 0.898 & 0.981 & 0.940 & 1.135 & 0.948 & 0.919\\
 &  & ( 0.182) & ( 0.219) & ( 0.339) & (-0.125) & ( 0.117) & ( 0.163)\\
 & M & 0.802 & 0.919 & 1.023 & 1.082 & 0.867 & 0.865\\
 &  & ( 0.039) & ( 0.278) & ( 0.979) & (-0.265) & ( 0.181) & ( 0.140)\\
 & L & 0.959 & \cellcolor{blue!25}1.374 & 1.193 & \cellcolor{blue!25}1.884 & 1.050 & \cellcolor{blue!25}1.286\\
 &  & (-0.188) & \cellcolor{blue!25}(-1.489) & ( 0.058) & \cellcolor{blue!25}(-2.660) & ( 0.070) & \cellcolor{blue!25}(-1.613)\\
\addlinespace
\multicolumn{8}{l}{\textbf{One-year-ahead}}\\
Homosk. & S & 0.950 & 1.007 & 0.978 & 1.243 & 1.045 & 1.039\\
 &  & ( 0.317) & ( 0.267) & ( 0.496) & (-0.155) & ( 0.107) & ( 0.071)\\
 & M & 0.988 & 0.984 & 0.979 & 1.067 & 1.020 & 1.036\\
 &  & ( 0.312) & ( 0.227) & ( 0.516) & (-0.147) & ( 0.159) & ( 0.048)\\
 & L & 0.969 & 0.973 & 0.997 & 0.990 & 0.953 & 0.954\\
 &   & ( 0.163) & (0.133) & ( 0.339) & ( 0.096) & ( 0.072) & ( 0.070)\\
\addlinespace
SV & S & 0.952 & 0.948 & 0.979 & 1.023 & 1.047 & 1.049\\
 &  & ( 0.076) & ( 0.171) & ( 0.118) & ( 0.105) & ( 0.125) & ( 0.106)\\
 & M & 0.951 & 0.950 & 0.999 & 0.995 & 1.005 & 1.026\\
 &  & ( 0.215) & (-0.036) & ( 0.473) & (-0.046) & ( 0.181) & ( 0.120)\\
 & L & 0.966 & \cellcolor{blue!25}1.149 & 1.024 & \cellcolor{blue!25}1.351 & 1.018 & \cellcolor{blue!25}1.292\\
 &  & (-0.013) & \cellcolor{blue!25}(-1.924) & ( 0.363) & \cellcolor{blue!25}(-2.198) & ( 0.120) & \cellcolor{blue!25}(-1.760)\\
\bottomrule
\end{tabular} 
}
\vspace{-.3cm}
\caption*{\footnotesize \textbf{Notes.} The table reports mean squared forecast errors relative to the large  BVAR with SV and average log predictive likelihood differences (in parentheses) between a given model and the large BVAR with SV. The BVAR-SV is estimated by setting $G=1$. The blue shaded cells are absolute MSEs and LPLs.  Bold numbers indicate the best performing model for a given variable and horizon.}
\end{threeparttable}
\end{table}

\begin{table}[ht!]
\caption{Tail Forecasting performance: 1977:Q1 to 2022:Q1}
\label{tab:tails}
\centering
\begin{threeparttable}
\scalebox{0.95}{
\begin{tabular}[t]{llrrrrrr}
\toprule
& &  \multicolumn{2}{c}{GDPC1} & \multicolumn{2}{c}{UNRATE} & \multicolumn{2}{c}{CPIAUCSL} \\
 & & DPM & $G=1$ & DPM & $G=1$ & DPM & $G=1$\\
\midrule
\multicolumn{8}{l}{\textbf{One-quarter-ahead}}\\
Homosk. & S & 0.882 & 0.890 & 1.295 & 1.705 & 0.919 & 0.931\\
 &  & (0.996) & (1.092) & (0.957) & (1.037) & (1.029) & (1.059)\\
 & M & 0.883 & 0.897 & 1.410 & 1.453 & 0.913 & 0.941\\
 &  & (1.038) & (1.016) & (1.085) & (0.974) & (0.943) & (0.942)\\
 & L & 0.944 & 0.948 & 1.436 & 1.261 & 0.941 & 0.944\\
 &  & (1.034) & (1.065) & (0.959) & (0.953) & (0.919) & (0.947)\\
\addlinespace
SV & S & 0.915 & 0.928 & 0.900 & 0.916 & 0.941 & 0.943\\
 &  & (0.832) & (0.881) & (0.966) & (1.035) & (1.004) & (0.945)\\
 & M & 0.941 & 0.914 & 1.462 & 0.940 & 0.942 & 0.973\\
 &  & (1.021) & (0.864) & (1.148) & (1.002) & (0.865) & (0.881)\\
 & L & 1.000 & \cellcolor{blue!25}0.183 & 1.573 & \cellcolor{blue!25}0.109 & 1.003 & \cellcolor{blue!25}0.211\\
 &  & (1.139) & \cellcolor{blue!25}(0.163) & (1.053) & \cellcolor{blue!25}(0.148) & (1.061) & \cellcolor{blue!25}(0.175)\\
\addlinespace
\multicolumn{8}{l}{\textbf{One-year-ahead}}\\
Homosk. & S & 0.817 & 0.836 & 0.919 & 0.963 & 1.028 & 1.038\\
 &  & (0.777) & (0.830) & (0.953) & (0.947) & (0.982) & (0.972)\\
 & M & 0.871 & 0.834 & 1.139 & 0.866 & 1.031 & 1.040\\
 &  & (0.865) & (0.805) & (1.143) & (0.940) & (0.951) & (0.953)\\
 & L & 0.887 & 0.890 & 0.897 & 0.896 & 1.006 & 1.012\\
 &  & (0.891) & (0.892) & (0.929) & (0.929) & (0.934) & (0.939)\\
\addlinespace
SV & S & 0.846 & 0.852 & 0.905 & 0.924 & 0.997 & 1.022\\
 &  & (0.817) & (0.844) & (0.955) & (1.036) & (0.957) & (0.991)\\
 & M & 0.888 & 0.861 & 1.211 & 0.944 & 1.021 & 0.997\\
 &  & (0.944) & (0.843) & (1.208) & (1.017) & (0.985) & (0.967)\\
 & L & 1.019 & \cellcolor{blue!25}0.216 & 1.179 & \cellcolor{blue!25}0.162 & 1.031 & \cellcolor{blue!25}0.204\\
 &  & (1.072) & \cellcolor{blue!25}(0.199) & (1.112) & \cellcolor{blue!25}(0.191) & (1.017) & \cellcolor{blue!25}(0.208)\\
\bottomrule
\end{tabular}
}
\vspace{-.1cm}
\caption*{\footnotesize \textbf{Notes.} The table reports relative quantile scores (10\% and 90 \%, in parentheses)  to the large  BVAR with SV.  The BVAR-SV is estimated by setting $G=1$. The blue shaded cells are absolute quantile scores. Bold numbers indicate the best performing model for a given variable and horizon.}
\end{threeparttable}
\end{table}

With some exceptions, it can be seen that DPM models perform well for all variables. The gains are particularly pronounced for point forecasts, where GDP growth and inflation forecasts are sometimes close to 20 percent more precise than the forecast produced by the benchmark model. The only variable where gains are muted (and sometimes even negative) is the unemployment rate. In this case, we sometimes find slight improvements in point forecasting accuracy but these never exceed 10 percent. 
When we consider  density forecasts we often find positive numbers in the parentheses. This indicates that the gains in terms of point forecasts also translate into more accurate density predictions.

To analyze whether the DPM component improves forecasts we can compare  each of the BVAR-DPM models with the equivalent model with $G=1$. For all three variables, two forecast horizons and for both forecast metrics, we almost always find the DPM version of the model to forecast better than the Gaussian one. In addition, versions of all models with SV tend to forecast better than homoskedastic versions of each model. This corroborate the finding in the forecasting literature that points towards the necessity for using flexible assumptions on the shocks to improve forecasts \citep{clark2011real, huber2019adaptive, carriero2022addressing}.

In terms of VAR dimension, it is interesting to note that the strongest performance for the DPM models arises with small and medium VARs. With large VARs, the evidence is more mixed both with regards to the need for DPM and with regards to the need for SV. We conjecture that in the large VAR, the explanatory power of the right hand side variables can mop up some (but not all) of the need to allow for non-Gaussianity or volatility change.

Table \ref{tab:tails} presents evidence on tail forecasting performance. The table shows 10\%  and 90\% QSs relative to the large VAR with SV so that numbers smaller than one indicate that a given model produces more accurate tail forecasts than the benchmark. 

We would expect the DPM to be particularly good in capturing tail behavior in a way that the Gaussian model cannot. And, with some exceptions, Table \ref{tab:tails} confirms this expectation. The pattern of results is similar to those found in Table \ref{tab: forecast_LPLMAE}, but are slightly stronger in favor of BVAR-DPM models. More specifically, we again find strong gains of the different models relative to the large BVAR with SV. Depending on the model size, we also find accuracy gains from using the DPM specification relative to setting $G=1$. These gains are more pronounced for  one-year-ahead tail forecasts of output growth and the unemployment rate and smaller-sized models.

\section{Conclusions}
In this paper we propose a new specification for the errors in a VAR which takes a particular additive form involving two components. The first is a homoskedastic error and a conventional inverse Wishart prior can be used for its covariance matrix. The second is a diagonal error covariance matrix with diagonal elements following SV processes. We show that, by adopting this additive form, we gain two major advantages. First, computation is much faster since it allows for equation-by-equation estimation. Second, posterior and predictive inference does not depend on the way the variables are ordered in the VAR. We then extend this model to allow the first error to follow a DPM. We discuss, both theoretically and empirically, the great flexibility that is obtained by doing so. In addition, we develop a computationally fast MCMC algorithm which allows for posterior and predictive inference in high dimensional non-parametric VARs. Our empirical results, using artificial and real data, show the flexibility of our approach. In our forecasting exercise, we find that the added flexibility of our BVAR-DPM in particularly useful in extreme times such as the financial crisis and the Covid-19 pandemic.

\clearpage


\clearpage
\small{\setstretch{0.85}
\addcontentsline{toc}{section}{References}
\bibliographystyle{cit_econometrica.bst}
\bibliography{lit}}
\newpage

\begin{appendices}
\renewcommand\thefigure{\thesection.\arabic{figure}}    

\section{Data Appendix}
\setcounter{table}{0}
\renewcommand{\thetable}{A\arabic{table}}
\begin{table}[h!]
\caption{Description of the Dataset} \label{tab: dataset}
\scalebox{0.65}{
\begin{tabular}{ZllccccZ} 
  \hline
   & FRED Code & Description & Transformation Codes & S & M & L & XL \\ 
  \hline
 & GDPC1 & Real Gross Domestic Product & 5 &                X& X & X & X \\ 
   & PCECC96 & Real Personal Consumption Expenditures & 5 &  &  & X & X \\ 
   & FPIx & Real private fixed investment  & 5 &  &  & X & X \\ 
   & GCEC1 & Real Government Consumption Expenditures and Gross Investment & 5 &  &  & X & X \\ 
   & INDPRO & IP:Total index Industrial Production Index (Index 2012=100) & 5 &  &  & X & X \\ 
   & CUMFNS & Capacity Utilization:  Manufacturing (SIC) (Percent of Capacity) & 1 &  &  & X & X \\ 
   & PAYEMS &  Emp:Nonfarm All Employees: Total nonfarm (Thousands of Persons) & 5 &  &  & X & X \\ 
   & CE16OV & Civilian Employment (Thousands of Persons) & 5 &  &  & X & X \\ 
   & UNRATE & Civilian Unemployment Rate (Percent) & 2 & X & X & X & X \\ 
     & AWHMAN & Average Weekly Hours of Production and Nonsupervisory Employees:  Manufacturing (Hours) & 1 &  &  & X & X \\ 
    & CES0600000007 & Average Weekly Hours of Production and Nonsupervisory Employees:  Goods-Producing & 2 &  &  & X & X \\ 
   & CLAIMSx & Initial Claims & 5 &  &  & X & X \\ 
     & GDPCTPI & Gross Domestic Product: Chain-type Price Index & 6 &  &  & X & X \\ 
    & CPIAUCSL & Consumer Price Index for All Urban Consumers:  All Items & 6 & X & X & X & X \\ 
     & PPIACO & Producer Price Index for All Commodities  & 6 &  &  & X & X \\ 
    & WPSID61 & Producer Price Index by Commodity Intermediate Materials:  Supplies \& Components & 6 &  &  & X & X \\ 
   & WPSID62 & Producer Price Index:  Crude Materials for Further Processing  & 6 &  &  & X & X \\ 
     & COMPRNFB & Nonfarm Business Sector:  Real Compensation Per Hour (Index 2012=100) & 5 &  &  & X & X \\ 
     & ULCNFB & Nonfarm Business Sector: Unit Labor Costs for All Workers & 5 &  &  & X & X \\ 
   & CES0600000008 & Average Hourly Earnings of Production and Nonsupervisory Employees: & 6 &  & X & X & X \\ 
   & FEDFUNDS & Effective Federal Funds Rate (Percent) & 2 & X & X & X & X \\ 
    & BAA10YM & Moody's Seasoned Baa Corporate Bond Yield Relative to Yield on 10-Year Treasury & 1 &  &  X& X & X \\ 
    & GS10TB3Mx & 10-Year Treasury Constant Maturity Minus 3-Month Treasury Bill, secondary market & 1 &  & X & X & X \\ 
   & CPF3MTB3Mx & 3-Month Commercial Paper Minus 3-Month Treasury Bill, secondary market & 1 &  &  & X & X \\ 
   & M2REAL & Real M2 Money Stock & 5 &  &  & X & X \\ 
   & BUSLOANSx & Real Commercial and Industrial Loans, All Commercial Banks & 5 &  &  & X & X \\ 
   & CONSUMERx & Real Consumer Loans at All Commercial Banks  & 5 &  &  & X & X \\ 
   & S.P.500 & S\&P's Common Stock Price Index:  Composite & 5 &  & X & X & X \\ 
   \hline
\end{tabular}
}
   \smallskip
\begin{minipage}{\linewidth}\small
\tiny \textbf{Notes}: This table provides an overview of the dataset employed. The transformation codes are applied to each time series $\bm Y_j$ and described in \cite{mccracken2020fred}: (1) no transformation; (2) $\Delta y_{jt}$; (3) $\Delta^2 y_{jt}$; (4) $\log (y_{jt})$; (5) $\Delta \log (y_{jt})$; (6) $\Delta^2 \log (y_{jt})$; (7) $\Delta (y_{jt}/y_{jt-1} - 1)$. 'X' marks the inclusion of a variable into one of the datasets.
\end{minipage}
\end{table}

\section{Additional empirical results}
\setcounter{figure}{0}    
\subsection{Impulse responses across clusters}

\begin{figure}[h!]
    \centering
    \begin{subfigure}{1.1\textwidth}
            \caption{GDPC1}
        \includegraphics[width=\textwidth]{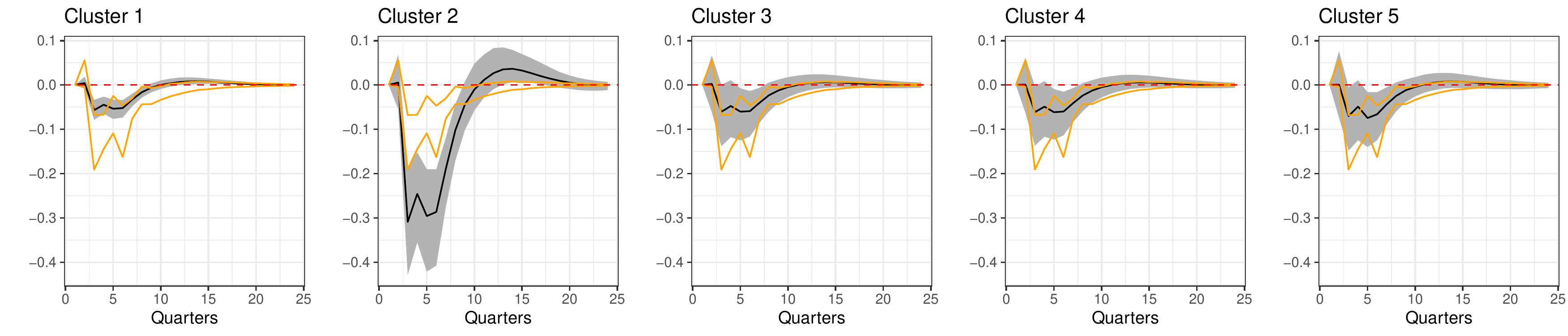}
        \label{fig:gdpc1}
    \end{subfigure}
        \begin{subfigure}{1.1\textwidth}
                \caption{CPIAUCSL}
        \includegraphics[width=\textwidth]{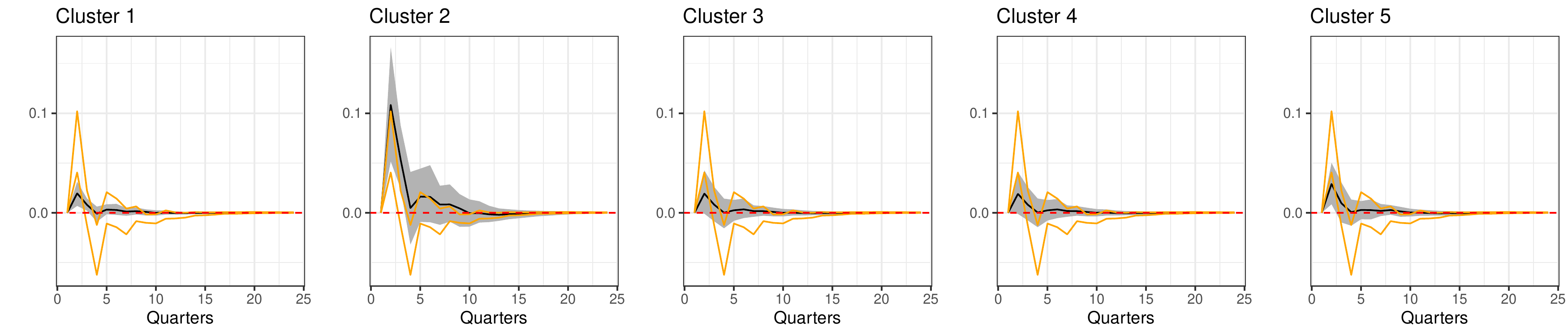}
        \label{fig:cpi}
    \end{subfigure}
    \begin{subfigure}{1.1\textwidth}
            \caption{UNRATE}
        \includegraphics[width=\textwidth]{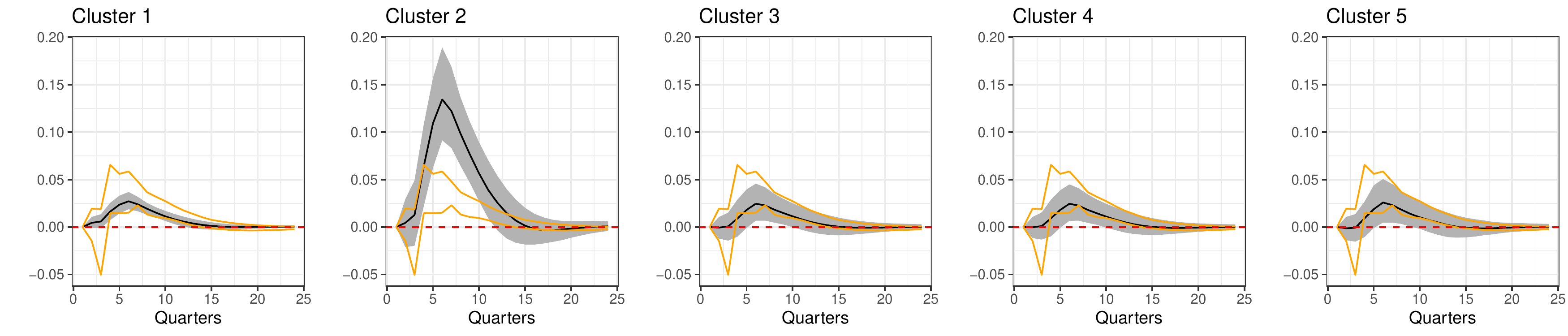}
        \label{fig:unrate}
    \end{subfigure}
        \begin{subfigure}{1.1\textwidth}
                \caption{FEDFUNDS}
        \includegraphics[width=\textwidth]{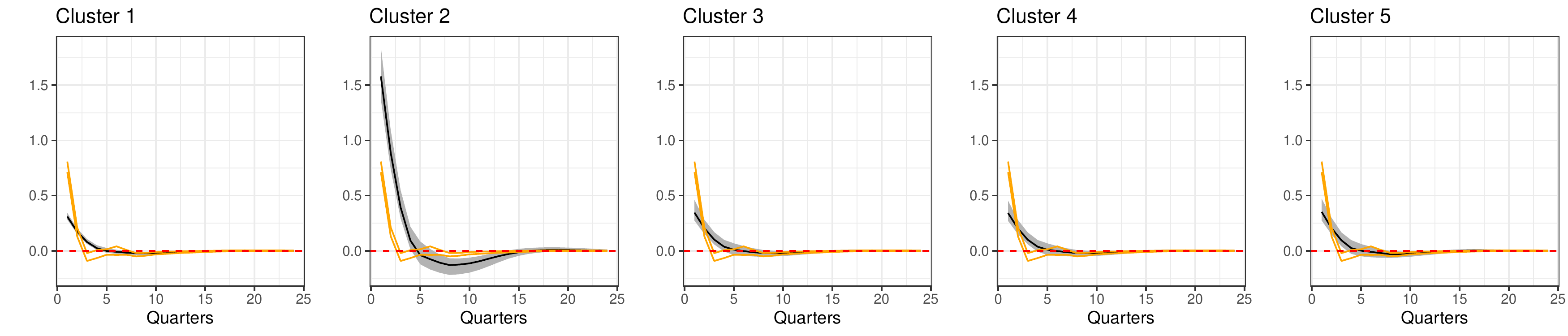}
        \label{fig:fedfunds}
    \end{subfigure}
    \caption*{\footnotesize \textbf{Notes}: The black solid line refers to the posterior mean IRF from the BVAR-DPM with SV across different clusters. The gray shaded area are the 16th/84th credible intervals of the BVAR-DPM while the orange lines refer to the 16th and 84th of the BVAR with $G=1$.}
\end{figure}

\begin{figure}[t!]\ContinuedFloat
    \centering
    \begin{subfigure}{1.1\textwidth}
            \caption{GS10}
        \includegraphics[width=\textwidth]{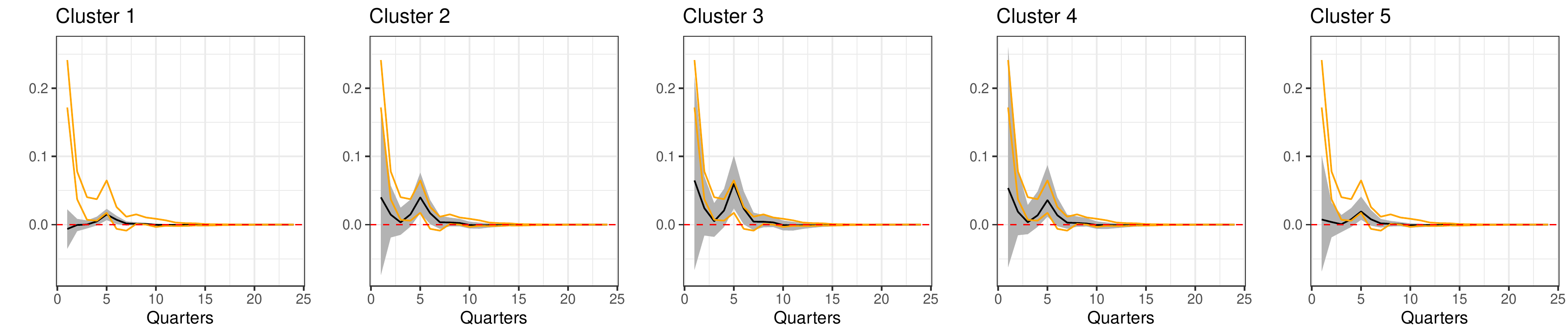}
        \label{fig:gs10}
    \end{subfigure}
        \begin{subfigure}{1.1\textwidth}
                \caption{S\&P 500}
        \includegraphics[width=\textwidth]{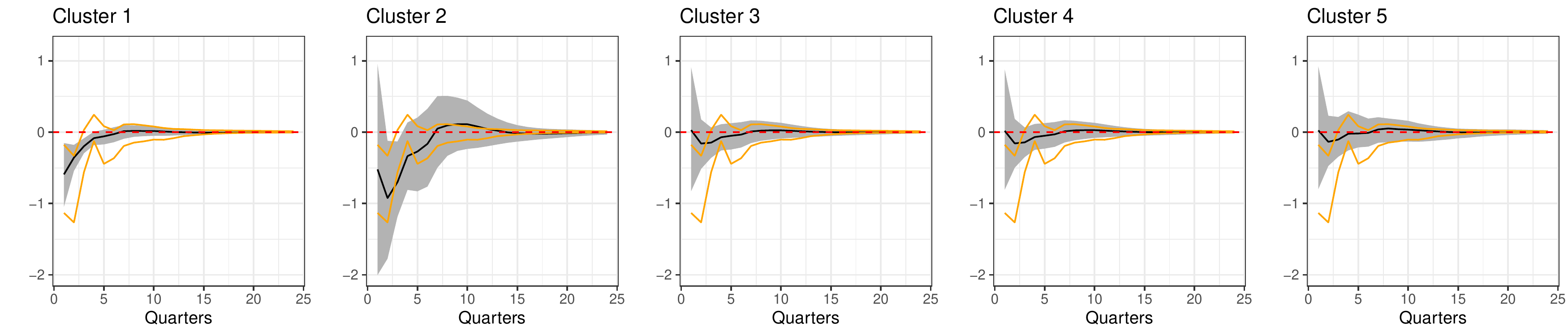}
        \label{fig:cpi}
    \end{subfigure}
    \caption*{\footnotesize \textbf{Notes}: The black solid line refers to the posterior mean IRF from the BVAR-DPM with SV across different clusters. The gray shaded area are the 16th/84th credible intervals of the BVAR-DPM while the orange lines refer to the 16th and 84th of the BVAR with $G=1$.}
    \caption{Impulse responses to a monetary policy shock for the different clusters.}
\end{figure}

\end{appendices}

\end{document}